\documentclass[onecollarge]{svjour2}                    
\smartqed  
\usepackage{graphicx}
\usepackage{latexsym, amssymb}
\DeclareRobustCommand{\schwa}{\rotatebox[origin=c]{180}{$e$}}
\begin{document}

\title{Determination of Dark Energy and Dark Matter from the values of Redshift for the present time, Planck and Trans-Planck epochs of the Big-Bang model
}


\author{Asger G. Gasanalizade         \and
        Ramin A. Hasanalizade 
}


\institute{Asger G. Gasanalizade \at
              Shemakha Astrophysical Observatory, National Academy of Sciences of the Azerbaijan Republic, AZ 5616, Shemakha, P.O. Pirkuli, Azerbaijan\\
              \email{gasanalizade@rambler.ru}           
           \and
           Ramin A. Hasanalizade \at
              Shemakha Astrophysical Observatory, National Academy of Sciences of the Azerbaijan Republic, AZ 5616, Shemakha, P.O. Pirkuli, Azerbaijan\\
              \email{hasanalizade@hotmail.com}    
}

\date{Received: date / Accepted: date}

\maketitle

\begin{abstract}
{As an alternative to the Standard $\Lambda CDM$ cosmology model in which the cosmological redshift quantified by relation $f(z) \sim (1+z)$, and presented the Universe Dark Energy $\Omega_{DE}$ as an Einstein's Cosmological Constant $\Lambda$, we have developed a new modified Freundlich's (quantum relativity) redshift (MFRS) mechanisms, which provide a precise solutions of the Dark Energy and Dark Matter problems. We apply the joint solution of three MFRS equations for concordances quantize bounce Planck hierarchy steps. Simultaneous scaling solutions of MFRS equations in logarithmic scale appropriate to three cosmological epoch's, yields a currently testable predictions regarding the Dark Matter $\Omega_{DM} = 0.25$, and Dark Energy $\Omega_{DE} = 0.75$. These predictions coincides with the recent observational data from WMAP and other a key supernovae SNe Ia findings. Thus, the presence of Dark Matter and Dark Energy had already been not only detected observationally, but also confirmed theoretically with the very compelling accuracy. From the WMAP7 and our predicted ages we find a value of the Hubble constant $H_0 = (65.6 \pm 0.6) km \cdotp s^{-1} Mpc^{-1}$ which is excellent agreement with the Planck 2013 results XVI. Compared with the ``holographic scenario'' results, we find an important coincidence between our new and ``holographic'' parameters. We discuss the connection hierarchy between the multiverse masses and examine the status of the cosmic acceleration. The product of the age of the Universe into the cosmic acceleration in each cosmological epochs --including present day are constant and precisely corresponds to an possible observable-geophysical parameter $g_U = 9.50005264_{265} (exact) \times (m/s^2)$. For the derived by WMAP7 age of the  Universe $t_{W7} = 13.75(13) \times 10^9 yr$,  we find the relevant acceleration $a_{W7} = 6.91(65) \times 10^{-10} m/s^2$. The predicted value of $t_0 = 9.0264_9(51) \times 10^2 Gyr$ is consistent with the background acceleration. $a_0 = 1.05246_4(61) \times 10^{-11} m/s^2$.}
\end{abstract}

\section{Introduction}
\label{intro}
The mysterious nature of Dark Energy (DE) is clearly on of the outstanding puzzles of the present quantum gravity and particle physics united in the quantum cosmology. In \textit {``Report of the Dark Energy Task Force (DETF)''} [3] the problem of understanding the dark energy is called out prominently in major policy documents such as the \textit {``Quantum Universe Report and Connecting Quarks with the Cosmos, and it is no surprise that it is featured as number one in Science magazine's list of the top ten Science problems of our time.''} Nowadays, among these science questions the nature of DE is identified as \textit {``probably the most vexing''} [3].

The First of DETF recommendation states: \textit {``We strongly recommend that there be an aggressive program to explore dark energy as fully as possible, since it challenges our understanding of fundamental physical laws and the nature of the cosmos''} [3]. 

The corresponding astrophysics measurements show that Universe is spatially flat, contains about (73--75)\% of the DE (or the near equivalent Einstein's cosmological constant $\Lambda$, see, i.e. [104]) and about (23--30)\% in the form Dark Matter (DM) (see e.g.  [11, 61, 122], and references therein). These data are compatible with the recent Wilkinson Microwave Anisotropy Probe (WMAP) observations (see e.g.  [64, 68, 69, 80], and references therein). There many several cosmological efforts were made to understanding the origin of DE and DM (for review, e.g. [35, 85, 86, 101, 117, 128], and references therein). 

In the preceding paper [53] we make a first attempt at such a study, by compiling quantum cosmological parameters taken for Today, Planck and Trans-Planck epochs of the Universe. In the present paper we anew give a comprehensive analysis for the exact solution of these challenges. We discuss their observational and experimental viewpoints, believe that exact explanations of the DM and DE problem lie outside of Standard model (SM) [52], since the present-day failures of the ``standard model of cosmology require a new''[74] reality. Thus, this paper is partially similar to the one of an Dark Matter and Dark Energy scenario, which we have considered previously [53], with the only difference that here the problem discussed in more detail by the adding the new knowledge and references. In the end of work [52] (see, also, [53]) it is mentioned that conception Synthesis of the Big Bang model with the modified Freundlich's redshift (MFRS) at Trans-Planck and the Planck scales provide a powerful tool to probe this problem. Further, based on combining shape constraints on simultaneous scaling solution of MFRS equations in logarithmic scale, a sequence of present day, Planck and Trans-Planck epochs cosmological parameters gives these ``puzzling parameters,'' rigorously defined as $\Omega_{DE} = 0.75$ and  $\Omega_{DM} = 0.25$.These predicted values of the Dark Energy and Dark Matter parameters are coincides with the predictions of the ``Millennium-I and II Simulations'' [18, 82] cosmological scenarios, and are compatible with the recent WMAP observational [64, 68, 69, 80] data. The supernova SNe Ia high redshift data (see e. g. [66, 69, 118]) was an ideal object for the application of this model. The calculation showed that results of validity of accelerated expansion of the Universe based on the scenario of flat Friedman-Robertson-Walker (FRW) model are highly esoteric. The model that opened the way for a simple interpretations of the SNe Ia high redshift data (see e. g.  [4, 83, 84]) perhaps is the $\Lambda$-Cold-Dark-Matter ($\Lambda$CDM) model with the some ``hot'' complementary. 

This paper is organized as follows: In Sec. 2 we summarize the necessary information about the discrete MFRS formalism under discussion. Here and what follows the subscripts \textit {``Pl''} and \textit {``TPl''} denotes values of Planck's and Trans-Planck's epoch parameters of MFRS equations, respectively. In Sec. 3 we choose a family of MFRS equations for the three steps discrete $10^{62}$ folding phase transition from the current epoch down to Trans-Planck's regime. Here along with the well-known Planck's parameters $m_{Pl}$, $T_{Pl}$, $l_{Pl}$ and $t_{Pl}$ [41, 110], it is given also some new quantum cosmological parameters. In Sec. 4 the detailed analysis of quantum cosmological parameters in the latest post Planck's epochs is given. Here we apply the logarithmic method to joint solution of these three cosmological redshift equations. Thereby, we firstly derive the corresponding predicted key parameters of $\Omega_{DM}$ and $\Omega_{DE}$ from the full combinations of above MFRS equations. In Sec. 5 we give refined version of state parameter of the Dark Energy w. In the next 6 Sec. we compare predicted values $\Omega_{DM}$ and $\Omega_{DE}$ with the observational data. The Sec. 7 is devoted to the age of Universe. The nature of Multiverse is discussed in Sec. 8. In Sec. 9 we briefly describe the latest Planck's epochs. In Sec. 10 the problem of accelerated expansion has been analyzed. In Sec. 11 the recommended values of Dark Energy and Dark Matter in physical units are presented. Finally, in the last Sec. 12 we draw our concluding remarks.

\section{Discrete Redshift Formalism and its Quantum Cosmological Parameters }
Traditionally, it is proposed that the Universe originated at $t \rightarrow 0$ by an event called the \textit {``Big Bang''} from a single point called a singularity [72]. As has already been noted, we firstly applied a new quantum cosmological representation for the creation of the Universe after the \textit {Big Bang}, based on the MFRS in some fixed steps of cosmological epochs [52, 53]. The basic idea in the beginning of this creation history of the Universe is that the time evolution between the birth and death for our permanently stepped expanding Universe divided into five distinct Plank epochs separated by steps $n = 2,1,0, - 1, -2$ at $z_{Pl}^n$. In this quantum cosmological scenario Dark Energy and Dark Matter can be predicted by simultaneous solutions of the first three Planck epochs of MFRS equations, which are relevant to the current observable Universe $(z_{Pl}^0 = z_0)$, Planck epoch $(z_{Pl})$, and Trans-Planck epochs $(z_{Pl}^2 = z_{TPl})$. (Moreover, by the Planck MFRS hierarchy presentation are reasons to assume that there available a further steps with the $n = 3, 4, \cdots, 8$) (These steps of MFRS are considered in Sec.3).

While ``the observable Universe exhibits several time asymmetries called of arrow of time'', in logarithmic scale these results does not asymmetry ``distinctions between past, present, and future'' [53]. Then, for all history of our observable Universe with the ``characteristic mass $M_U$'' [52, 53, 150] after of Big Bang up to a Big Rip (the end of our Universe), for expanding Universe we assume five discrete steps (eras) of fundamental MFRS equations with $z^n$. Eventually, the beginning and break down of our Universe may then have been at some MFRS between $z_{Pl}^2 \sim 10^{125}$ and $z_{Pl}^{-2} \sim 10^{-126}$. As a result, for the present-day observable Universe we have a first cosmological MFRS by $z_{Pl}^0 = z_0 = 1$ at the age of $t_0 \sim 10^{19} s$ (the expanding time from the Big Bang to $z_0 = 1$) and predicted constant cosmic background radiation (CBR) temperature of $T_0(CBR) \sim 2.766_2 K$, (like to the cosmic microwave background (CMB) temperature $T_0(CMB) = 2.725_2 K$ measured COBE in 1992)\footnote {This value correspond to predicted temperature $T_0(CBR)$ is calculated completely on the basis of fundamental physical constants of $c, \hbar, k, G$ and Freundlich-Melvin constants $A_S$. Clearly, parameter $T_0(CBR)$ can expect as a quantum temperature of the novel quantum cosmology.}. (Here and in the following, the subscript ``0'' denotes the present epoch values). In second, we assumed that the Planck epoch MFRS (Planck redshift) by $z_{Pl} \sim 10^{62}$ at the Planck time of $t_{Pl} \approx 5 \times 10^{-44} s$ with temperature of $T_{Pl}^{'} \sim 10^{32} K$ is landmark, in accord with Planck's concept [110]. The creation of a homogenous Universe in the Trans-Planck ``epoch'' can be described by a Trans-Planck MFRS (Trans-Planck redshift) equation by $z_{TPl} \sim 10^{125}$, at time of $t_{TPl} \sim 10^{-106} s$, and constrained with temperature as $T_{TPl} \sim 10^{63} K$. This step proposed that the present-day Universe begin from a non-zero radius $l_{TPl} \sim 10^{-98} m$, with a total energy of $M_U \cdot c^2 \sim 10^{81} GeV$. So, these quantum gravity parameters correspond to the creation of our observable Universe from the Big Bang, which are derived without violate the Heisenberg's uncertainty principle [52]. However, here we convinced\footnote {In this work it has been mistakenly argued that the creation moment of the Universe (the Big Bang) begin at the $t_{OTPl} \sim 10^{-169} s$ (take Over Trans-Planck energy scales). Presented in subsection 5.3.4 equations are valid in the first order approximations by relation of $\schwa = t_{OTPl} \cdot T_{OTPl}^2$, where $T_{OTPl} \sim 10^{94} K$. Initially this is true. However, when time became $t = t_{OTPl}$ the quantum relativistic approximation $(M_U \cdot c^2) \cdot t_{OTPl} = \hbar$ will do not and numerical simulations by the Eqs. (5.20a)--(5. 21a) may be exists before creation of the today Universe. But that was before basic MFRS $z_{Pl}^3$ was established: here exists of the quantum relation $(M_{PMV} \cdot c^2) \cdot t_{OTPl} = \hbar$ evolving for the Planck Multiverse $M_{PMV}$ was yet to come (see Subsection 3.4).} that the generalized uncertainty principle is not generally valid at time $t \sim 10^{-169} s$. In addition, the moment $t_{TPl}$ is along way from initial zero singularity when in the Friedman universe $t \rightarrow 0$, and the energy density was infinity, i.e. $u_v(t) \rightarrow \infty$. Thus, this ultimate historical difficulty of singularity in Friedman cosmology model is eliminated. 

Notice also that in the pre-current evolutions of the Universe the MFRS is $z > 1$, and in post- current this is $z < 1$, respectively. The value $z \equiv z_0 \equiv1$ corresponds to the current Universe.

\section{The MFRS Equations for Three Discrete Steps of Planck's Epochs}
When MFRS model was proposed in 2006 [51] it contained three types of MFRS equations, covering the present, electromagnetic and Planck's epochs, with utilizing $z = z_0 = 1$, $z = z_e \sim 2 \times 10^{40}$, and $z = z_{Pl} \sim 5 \times 10^{62}$, respectively. In subsequently, in detailed description of a MFRS models [52,53] it has been argued that the current values of $\Omega_{DE}$ and $\Omega_{DM}$ can be determined from combinations of the several bounce steps of Planck's MFRS equations. In this sense, below we anew use three complex set of discrete steps $z_{Pl}^n$ Planck's MFRS equations from $z \equiv z_{Pl}^0 \equiv z_0$ at $n \equiv 0$ (Today), to $z_{Pl}^1 \equiv z_{Pl}$ at $n \equiv 1$ (Planck) and $z_{Pl}^2 \equiv z_{TPl}$ at $n = 2$ (Trans-Planck). As illustrated in Section 4 the most plausible estimation of the Dark Matter (making of 0.25 parts) and the Dark Energy (making of 0.75 parts) of the present observable Universe energy, come from combining solutions of these three scaling steps of discrete Planck's MFRS equations. 
\label{sec:1}
\subsection{The Present-day Universe MFRS equation }
According to a quantum relativity approaches [51-53] the first bounce steps of MFRS for the our observable Universe may be \textit {``scaling scenario''} (for review, see [28], and references therein), for $n = 0$ characterized by the ratios
\begin{equation}
z_0 = \left( \frac {T} {T_0} \right)^2 = \frac {t_0} {t} = \frac {l_0} {l} = \frac {M_U} {m} = \cdots = \frac {a} {a_0} = \cdots \equiv (\schwa \cdot c \cdot A_S) \cdot T_0^2 =1 \mbox {,}
\end{equation}
where the $T_0$, $t_0$, $l_0$, $M_U$ and $a_0$ are the predicted key quantum relativity cosmological parameters for the cosmic background radiation (CBR) temperature, age of the Universe, horizon size comparable to the radius of curvature, the total mass and the background acceleration\footnote {The classical description $a_0$ is the Newtonian gravitational acceleration.} at the present Universe, since the Big Bang. Basically, these key constant scaling parameters for describing of our observable Universe are expressed in terms of the speed of light in vacuum $c$, \textit {the new quantum gravity-cosmological constant} $\schwa$ parameterized as $T_i^2 \cdot t_i$ [52, 53], the Newton's gravitational constant $G$, and the \textit {Freundlich-Melvin constant} $A_S \approx 2 \times 10^{-30} m^{-1} K^{-4}$ [46, 93]. In fact, one of the basic result follows from relation (1) is identity combinations today horizon $l_0$ and age of the Universe $t_0$ with the CBR temperature $T_0$. The relation between $T_0$ and $A_S$ is inextricable coupled to the relation between $T$, $t$ and constant $\schwa$, presenting in the general case, an intractable problem. This can, however, be solved in closed form under the following [21] assumption $A_S \cdot T_0^4 = 1 / R_U$, where $R_U$ is the radius universe. To determine the present \textit {``relict temperature''} $T_0$ we set $R_U = l_0$ in the above equation, finding
\begin{equation}
(l_0 \cdot T_0^4)^{-1} = A_S \mbox {, and } t_0 \cdot T_0^2 = \schwa \mbox {.}
\end{equation}
In this way [at MFRS of $z_0 =1$ based on (1)], we can predict the following present-epoch values\footnote {It is evident from Eq. (3), that the Universe may be accepted the negative temperature, which comes from the fact that the CBR temperature $T_0$ follows from $T_0^2 = (\schwa \cdot c \cdot A_S)^{-1}$.}  of the some today quantum-cosmological parameters as
\begin{equation}
T_0 = \pm(c \schwa A_S)^{-1/2} \mbox {, } t_0 = c \schwa^2 A_S \mbox {, } l_0 = (c \schwa)^2 A_S \mbox { and } a_0 = (\schwa^2 A_S)^{-1} \mbox {,}
\end{equation}
where the constant $\schwa$ is measured in $s \cdot K^2$ units. The constant $\schwa$ was the firstly proposed in 2004 [50]. Actually, for all cosmological epochs constant $\schwa$ can be written as [51-53]
\begin{equation}
\schwa \equiv \cdots \equiv t_0 \cdot T_0^2 \equiv \cdots \equiv t_{Pl} \cdot T_{Pl}^2 \equiv \cdots \equiv b(c_2 / c_1)E_{Pl} \mbox {,}
\end{equation}
where $b$ is Wien's displacement law constant, $c_1$ and $c_2$ are the first and second radiation constants, respectively and $E_{Pl}$ is the Planck energy. We already discussed the importance universality of quantum cosmological constant $\schwa$ in an above our works. Thus, proposed relationship by (4) does not vary in various epochs of the Universe. Then in either case, the set $(T_0, t_0)$, $(T_{Pl}, t_{Pl})$ and $(T_{TPl}, t_{TPl})$ pairs are enough to assure the robustness determine, which remains constant in time intervals $10^{-106} s \le t \lesssim 10^{82} s$ since the Big Bang. Because $\schwa$ is the \textit {universal quantum gravity-cosmological constant}, it may be of interest to include the quantized state $n = -1, -2$ into the computation, to obtain a realistic estimation in the final fate of the Universe history. In the latest post-current epochs [142] the constant $\schwa$ appear also at $z_{Pl}^{-1}$ and $z_{Pl}^{-2}$ RMFS equations ranging up to Big Rip. This problem is considered in Sec. 4 as $\schwa = T_i^2 \cdot t_i \cong 2.18 \times 10^{20} K^2 s$.

\label{sec:1}
\subsubsection{The Value $\schwa$ on the basis of Fundamental Physical Constants }
The value of the quantum-cosmological constant $\schwa$ cannot be calculated directly since the $T_0(CBR)$ temperature, and present age of the observable Universe $t_0$, or the Planck's temperature $T_{Pl}$ and Planck's time $t_{Pl}$ has not been determined. Here the basic idea, consistent with the MFRS is that, \textit {this new constant has the fundamental importance for application in all epochs of the expansion history of the Universe, i.e. it does not depend on age of the Universe}. With these assumptions there may be exception for an independent direct determination of constant $\schwa$ on the basis of current Fundamental Physical Constants (FPC). Therefore, we hopes that this constant in the near time would be included to CODATA list as one \textit {new quantum gravity- cosmological constant}.

Using the FPC of $b, c_1, c_2$  and $E_{Pl}$, from CODATA-2006 [95] early we obtain (see paper [52])
\begin{equation}
\schwa = 2.17955_5 (13) \times 10^{20} K^2 s \mbox {.}
\end{equation}

In this paper, we improve the numerical value of $\schwa$ by using recent CODATA-2010 data sets for the FPC [96]. In that case, the universal constant $\schwa$ could be defined as 
\begin{equation}
\schwa = b(c_2 / c_1) \cdot E_{Pl} = \frac {(\hbar / k^2) \cdot E_{Pl}} {4.965114231} = 2.179627_{6} (130) \times 10^{20} K^2 s \mbox {,}
\end{equation}
where $\hbar = 6.58211928(15) \times 10^{-16} eV \cdot s$ is Planck's constant, $k = 8.6173324(78) \times 10^{-5} eV \cdot K^{-1}$ is Boltzmann's constant, and $E_{Pl} = m_{Pl} \cdot c^2 = 1.220932(73) \times 10^{19} GeV$ is the Planck energy. 

However, in that background, fine adjustment of the constant $\schwa$ and other constants are limited by the accuracy of Newtonian constant of gravitation $G$, and the Stefan-Boltzmann constant $\sigma$. A useful picture for improve our calculations of the $\schwa$ is based on absolutely precision of the product of the $G \schwa^2$ and $\sigma$ given by
\begin{equation}
(G \schwa^2) \cdot \sigma = (4.965114231)^{-2} (\pi^2 / 60) \cdot c^3 = 1.797844453_{2} \times 10^{23} (m / s)^3 (exact) \mbox {.}
\end{equation}

The importance of this equation comes from the fact that the speed of light in vacuum $c$ in right side is an \textit {``exact''} and independent of any quantum-electromagnetic parameters. So if we adopt CODATA-2010 [96] values of $G = 6.67384(80) \times 10^{-11} m^3 kg^{-1} s^{-2}$ and $\sigma = (\pi^2 / 60) k^4 / \hbar^3 c^2 = 5.670373(21) \times 10^{-8} W \cdot m^{-2} K^{-4}$, then the constant $\schwa$ from the Eq. (6) can be anew directly calculated within the accuracy of $0.00006$, and equal
\begin{equation}
\schwa = 2.179627_{9} (130) \times 10^{20} K^2 s \mbox {.}
\end{equation}

Then, according to the estimates of Eqs. (6) and (8) today final revised result for \textit {the quantum gravity- cosmological constant} $\schwa$ is given by
\begin{equation}
\schwa = 2.179627_6 (130) \times 10^{20} K^2 s \mbox {.}
\end{equation}

In principle, by using the revised Eq. (7), the Newton's gravitational constant $G$ becomes calculable in terms of FPC $c, \hbar, k, \sigma$ and\textit {new quantum gravity-cosmological constant} $\schwa$.

On the other hand it is evident that, \textit {``the expanding universe is an excellent laboratory to study the effect of scale change on physical laws and physical constants''} [97]. After all, the derived new constant $\schwa$ offers way for MFRS equations in at all epochs of expanding Universe. These epochs corresponded to a quantized interval of $z_{Pl}^8 - z_{pl}^{-2}$, from the pre-Big Bang to the ``Big Rip'' [23]. Consequently, the constant $\schwa$ attained the status of \textit {fundamental quantum gravity- cosmological constant} for all cosmological epochs. The application of the new constant is very similar to the application of the Boltzmann constant $k$, which can be used especially at some new correct quantum-cosmological computations. 

\label{sec:2}
\subsubsection{The Mass of the Our Observable Universe}
Throughout this paper we introduce the total rest mass of the observable Universe $M_U$, accepted in Einstein theory as a \textit {``characteristic mass scale"} [150]. In addition, the quantum cosmological quantity this mass $M_U$ is defined as (See, also [52, 53]) 
\begin{eqnarray}
M_U \equiv \frac {c^2} {G} l_0 \equiv \frac {\schwa c^3} {G T_0^2}\equiv \frac {(\schwa c^2)^2} {G} A_S \equiv \frac {T_{TPl}} {T_0} m_{Pl} \equiv \cdots \nonumber \\  
= 1.15000_5(76) \times 10^{55} kg = 6.4510_6(21) \times 10^{81} GeV / c^2 \mbox {.}
\end{eqnarray}

It turns out that the mass $M_U$ is about $\sim 10^{25}$ times the mass $M_S$ of the Sun, that is, $M_S = 1.988435(27) \times 10^{30} kg$ [57]. Here and in what follows, numerical values of all physical constants and new cosmological parameters with the exception of constant $A_S$, are determined according to CODATA-2010 data [96]. However, it is worth noting that there difficult to asses the error in value of constant $A_S$. The accuracy $A_S$ was determined as follows: Our predicted cosmic (frequency independent) background radiation temperature given by $T_0(CBR) = (c \schwa A_S)^{-1/2} = 2.7662_1 K$ [52], comparable with the Cosmic Microwave Background $(CMB)$ black-body temperature $T_0(CMB) =2.725(1) K$ [44] (measured locally at redshift $z = 0$) and coincides with the $T_0(CMB) = (2.766 \pm 0.160) K$ derived at an $8.3 GHz$ by the Absolute Radiometer for Cosmology, Astrophysics and Diffuse Emission (\textit {ARCADE} 2) team measurements [121] (though this may be interpreted as another radiation not associated with $CMB$ temperature). However, the problem theoretical explanation for the actual interpretation of $CMB$ gets a worse when we come to the early Universe with the Planck time scale $t_{pl}$, or temperatures well beyond the Planck temperatur $T_{Pl}^{'}$ in Big Bang scenario [52], though this claim is questionable [143]. As is repeatedly mentioned, there is no underlying fundamental physical theory for the Trans-Planck regime [19, 28, 67, 94]. Since the constant $\schwa$ depends exclusively on FPC, then this can be used in all cosmological epochs. A possible constraint coming from the $CMB$ was discussed in [44, 121] (see, also, [10, 22, 40, 92]). Nonetheless, from comparison results of these temperatures, at once we estimate refined value of \textit {Freundlich-Melvin constant} [52] as 
\begin{equation}
A_S = 2.0000_0 (46) \times 10^{-30} m^{-1}  K^{-4} \mbox {.}
\end{equation}

Then the final values of above predicted quantum cosmological parameters $t_0, l_0, a_0$ and others may be estimated many times better than a few percents [52, 53, 60]. The numerical examples for the some current Universe parameters, at constant value of velocity of light in vacuum $c$ (here and further), are shown in Table 1  as a function of $T_0^2 (CBR)$. 
\begin{table}
\caption {The present-day quantum gravity cosmological parameter sets of the Universe.}
\renewcommand{\arraystretch}{1.8}
\begin{tabular}{llll}
\hline
Parameters & Symbol & Definition & Value\\
\hline
\textbf {Background temperature} & $T_0 (CBR)$ & $(\schwa \cdot c \cdot A_S)^{-1/2}$ & $2.7662_{05} (160) K$\\
\textbf {Modified Freundlich's Red Shift} & $z_0$ & $(\schwa \cdot c \cdot A_S) \cdot T_0^2$ & $1$\\
Present-day age of the Universe & $t_0 = T_U$ & $\schwa / T_0^2$ & $2.8484_{81} (16) \times 10^{19} s$\\
Horizon size & ${l_0} = c{t_0}$ & $\schwa \cdot c/T_0^2$ & ${8.5395_{32}}(49) \times {10^{27}}m$\\
Background acceleration &${a_0}$ & $(c/\schwa ) \cdot T_0^2$ & ${1.05246_{42}}(61) \times {10^{ - 11}}m/{s^2}$\\
 $G \times $Mass of the Universe & $G{M_U}$ & $(\schwa  \cdot {c^3}/T_0^2)$ & ${7.6749_{49}}(33) \times {10^{44}}{m^3}/{s^2}$\\
Present Cosmological constant &${\Lambda _0} = l_0^{ - 2}$ & ${(T_0^2/\schwa  \cdot c)^2}$ & ${1.3713_{00}}(16) \times {10^{ - 56}}{m^{ - 2}}$\\
Mass of the Graviton & ${m_{Gr}}$ & $(\hbar /\schwa \cdot {c^2})T_0^2$ & ${4.1192_{80}}(10) \times {10^{ - 71}}kg$\\
Present matter density & ${\rho _\Lambda }({t_0})$ & ${(T_0^2/\schwa )^2}/8\pi G$ & ${7.3478_{04}}(43) \times {10^{ - 31}}kg/{m^3}$\\
Present energy density & ${u_\nu }({t_0})$ & ${\rho _\nu }({t_0}) \cdot {c^2}$ & ${6.6038_{77}}(40) \times {10^{ - 14}}J/{m^3}$\\
Expanding velocity of the Universe & ${l_0}/{t_0} \equiv {a_0} \cdot {t_0}$ & $c$ & ${299792458_2}(exact)m/s$\\
Cosmological acceleration & $c/1yr$ & ${g_U}$ & ${9.50005264_{265}}(exact)m/{s^2}$\\
\hline
\end{tabular}
{\\
\\(Throughout this paper the figures in parentheses after the values give the one-standard-deviation \\
uncertainties in the last significant digits).}
\end{table}

\label{sec:3}
\subsubsection{Equality between density and pressure}
From Table 1 and Eq. (1) it is immediate that
\begin{equation}
{u_\nu }({t_0}) = {\rho _\nu }({t_0}) \cdot {c^2} = {6.6038_{8}}(40) \times {10^{ - 14}}J/{m^3}
\end{equation}

Recall, derived in SI units ${J/m^3}$ is named Pascal (Pa) and use for pressure measurements. 

An alternative view of this result is the identification of positive sign of the energy density ${u_\nu }({t_0})$ with a negative pressure parameter ${P_\nu }({t_0})$ taking into account that (see e.g. [109]) ${u_\nu }({t_0}) =  - {P_\nu }({t_0})$.

In this case, considering that 1Pa$ \approx {10^{ - 5}}$atm, the above negative pressure may be written also as
\begin{equation}
{P_\nu }({t_0}) = {6.6038_8}(40) \times {10^{ - 14}}Pa \approx {6.6038_8}(40) \times {10^{ - 21}}atm, 
\end{equation}
which confirmed the equation of state of the dark energy ${w_\nu }(t)$ for a flat Universe, and should be negative. In such (vacuum dominated) flat Universe case for the present-day value of the ${w_\nu }({t_0})$ correspond to 
\begin{equation}
{w_\nu }({t_0}) =  - {P_\nu }({t_0})/{\rho _\nu }({t_0}) =  - [a_0^2/8\pi G]/{u_\nu }({t_0}) =  - 1.00.
\end{equation}

\label{sec:2}
\subsection{The Planck's epoch MFRS equation }
In the most cosmological models the Planck parameters ${m_{Pl}}$, ${t_{Pl}}$, ${l_{Pl}}$and ${T_{Pl}}$ are accepted as a rigorously derived fundamental quantum gravity constants of the cosmology (see e.g. [75]). Except the Planck's temperature${T_{Pl}}$, these parameters somewhat already known (e.g. [41, 52, 53, 134]). By regarding Eq. (4) we first notice the fact that there is no agreement between ${T_{Pl}}$ and other Planck units [52] by the \textit {``scaling scenario''}.

As was determined in our previous works [52, 53] the MFRS for the Planck epoch step can be expressed by a \textit  {scaling relation} between Planck's parameters and the predicted present quantum relativity parameters at $n = 1$, leading to
\begin{eqnarray}
z_{Pl} \equiv N_{Pl} \equiv \left( \frac {T_{Pl}^{'}}{T_0} \right) ^2 \equiv \frac{t_0}{t_{Pl}} \equiv \frac {l_0}{l_{Pl}} \equiv \frac{{{a_{Pl}}}}{{{a_0}}} \equiv  \cdots  \equiv \frac{{{u_\nu }({t_{Pl}})}}{{{u_\nu }({t_0})}} \equiv \frac{{{\rho _\nu }({t_{Pl}})}}{{{\rho _\nu }({t_0})}} \equiv  \cdots  \nonumber \\
\equiv \frac{{{M_{PMV}}}}{{{M_U}}} \equiv \frac{{{M_U}}}{{{m_{Pl}}}} \equiv \frac{{{m_{Pl}}}}{{{m_{Gr}}}} \equiv  \cdots  = {5.2837_1}(31) \times {10^{62}},
\end{eqnarray}
where statistical dual ${N_{Pl}}$ coincides with redshift ${z_{Pl}}$ and equal to a number of Planck's particles at moment ${t_{Pl}}$. Here ${t_{Pl}} = {(\hbar G/{c^5})^{1/2}} = 5.39106(32) \times {10^{ - 44}}s$, ${l_{Pl}} = {(\hbar G/{c^3})^{1/2}} = 1.616199(97) \times {10^{ - 35}}m$ and ${m_{Pl}} = {(\hbar c/G)^{1/2}} = 2.17651(13) \times {10^{ - 8}}kg$ [96] are the Planck's time, length and mass, respectively. In particular, according to the \textit {scaling scenario} of Eq. (15) the reduced Planck's temperature $T_{Pl}^{'}$ previously determined in (see, [52]) here corrected, namely,
\begin{equation}
T_{Pl}^{'} = {(\schwa /{t_{Pl}})^{1/2}} = {T_{Pl}}/{(4.965114231)^{1/2}} = {6.358490_9}(23) \times {10^{31}}K,
\end{equation}
where ${T_{Pl}}$ is Planck's temperature [95, 96] defining as
\begin{equation}
{T_{Pl}} = {(\hbar {c^5}/G{k^2})^{1/2}} = 1.416833(85) \times {10^{32}}K.
\end{equation}

In this case $T_{Pl}^{'}$ is a fixed \textit {scaling} approach determination of Planck's temperature.

The Planck matter density within the time ${t_{Pl}}$ is given by
\begin{equation}
{\rho _\nu }({t_{Pl}}) = {3.8823_{7}}(23) \times {10^{32}}kg/{m^3},
\end{equation}
and the Planck energy density will be
\begin{equation}
{u_\nu }({t_{Pl}}) = {\rho _\nu }({t_{Pl}}){c^2} = {3.4893_0}(20) \times {10^{49}}J/{m^3}.
\end{equation}

Then, \textit {Planck's redshift is the quantum cosmological scaling ratio of present day parameters \\ $[T_0^{ - 2},{t_0},{l_0},{M_U},{a_0},{\rho _\nu }({t_0}),{u_\nu }({t_0}), \cdots ]$ of the our observable Universe to the Planck's parameters \\  $[T_{Pl}^{ - 2},{t_{Pl}},{l_{Pl}},{m_{Pl}},{a_{Pl}},{\rho _\nu }({t_{Pl}}),{u_\nu }({t_{Pl}}), \cdots ]$, and equal to ${z_{Pl}} \equiv {N_{Pl}}$.} 

It should noted that, the above coincident resembles ratio equality between present-day values of the radius, age and mass of the observable Universe and Planck parameters for length, time and mass in other form is roughly ($\approx 10^{61}$) proposed also by J. Casado. [27] 

\label{sec:3}
\subsection{The Trans-Planck's epoch MFRS equation }
Extensions of the MFRS model below to time interval with $t <  <  < {t_{Pl}}$, similar to above Planck's equation, allowed the Trans-Planck's epoch (TPl) following the Big Bang [Because we don't have reliable theory of Quantum gravity for earlier ${t_{Pl}}$ time (see, i.e. [34, 63, 119]), we are now able to probe the Trans-Planck's time as ${t_{TPl}} \equiv t_{Pl}^2$)]. The application of the MFRS to the Trans-Planck ``epoch'' is very similar to the application this in the Planck's epoch of the observable Universe). We show that inserting $n = 2$ in place $n = 0$ into (1) the Trans-Planck's MFRS is given by [52,53],
\begin{eqnarray}
{z_{TPl}} \equiv z_{Pl}^2 \equiv {N_{TPl}} \equiv {(\frac{{{T_{TPl}}}}{{{T_0}}})^2} \equiv \frac{{{t_0}}}{{{t_{TPl}}}} \equiv \frac{{{l_0}}}{{{l_{TPl}}}} \equiv \frac{{{a_{TPl}}}}{{{a_0}}} \equiv \frac{{{u_\nu }({t_{TPl}})}}{{{u_\nu }({t_0})}} \equiv \frac{{{\rho _\nu }({t_{TPl}})}}{{{\rho _\nu }({t_0})}} \equiv  \cdots  \equiv \nonumber \\
\equiv \frac{{{M_U}}}{{{m_{Gr}}}} \equiv {(\frac{{{M_U}}}{{{m_{Pl}}}})^2} \equiv \frac{{{M_{OTMV}}}}{{{M_U}}} \equiv \frac{{{m_{Pl}}}}{{{m_{Str}}}} \equiv  \cdots  = {2.7917_6}(30) \times {10^{125}},
\end{eqnarray}
where dual with the ${z_{TPl}} \equiv z_{Pl}^2$ parameter${N_{TPl}}$ is the number of created (graviton) particles in Trans-Planck epoch, and time ${t_{TPl}}$ is moment of the Big Bang. Then, the other parameters with the \textit {``TPl''} in the very early universe are still too limited to indicate the properties of the Big Bang. The analysis Eq. (20) shows us that the cosmological parameters of the Trans-Planck's MFRS are determined from the combinations of the present day and Planck's epochs MFRS parameters by the following relations
\begin{eqnarray}
{z_{TPl}} \equiv z_{Pl}^2/{z_0} \equiv {({M_U}/{m_{Pl}})^2} \equiv z_{Pl}^2 = {2.7917_{6}} \times {10^{125}},\\
{T_{TPl}} \equiv {(T_{Pl}^{'})^2}/{T_0} = {1.4615_{8}}(8) \times {10^{63}}K,\\
{t_{TPl}} \equiv t_{Pl}^2/{t_0} \equiv \hbar /{M_U}{c^2} = {1.0506_{6}}(6) \times {10^{ - 106}}s,\\
{l_{TPl}} \equiv l_{Pl}^2/{l_0} \equiv \hbar /{M_U}c = {3.0588_{3}}(18) \times {10^{ - 98}}m, \\
{m_{Gr}} \equiv m_{Pl}^2/{M_U} = {4.1192_{82}}(10) \times {10^{ - 71}}kg = {2.3107_{5}}(13) \times {10^{ - 35}}eV/{c^2}\\
{m_{Str}} \equiv m_{Gr}^2/{m_{Pl}} = {7.7961_{87}}(45) \times {10^{ - 134}}kg = {4.3733_{5}}(25) \times {10^{ - 98}}eV/{c^2}.\\
{\rho _\nu }({t_{TPl}}) \equiv \rho _\nu ^2({t_{Pl}})/{\rho _\nu }({t_0}) = {2.0513_{3}}(41) \times {10^{96}}kg/{m^3},\\
{u_\nu }({t_{TPl}}) \equiv u_\nu ^2({t_{Pl}})/{u_\nu }({t_0}) = {1.9625_{4}}(19) \times {10^{112}}J/{m^3},\\
{a_{TPl}} \equiv a_{Pl}^2/{a_0} = {2.9382_{3}}(17) \times {10^{114}}m/{s^2}.
\end{eqnarray}

The one crucial point of these developments is that the today Quantum cosmological constant ${\Lambda _0}$ is linked with the Planck and Trans-Planck's redshifts (for a more confirmative view, see also [52], Sect. 4) by relations
\begin{equation}
{({\Lambda _{0 \cdot }} \cdot l_{Pl}^2)^{ - 1}} \equiv {({l_0}/{l_{Pl}})^2} \equiv  \cdots  \equiv z_{Pl}^2 \equiv {z_{TPl}} \equiv N_{Pl}^2 \equiv {N_{TPl}}.
\end{equation}

Then, above sets of predicted parameters can be considered as one of most remarkable successes of \textit {new modified quantum gravity cosmology}, that led to the direct predictions birth parameters of the Universe, corresponding to a extremely tiny Trans-Planck time ${t_{TPl}}$ and short Trans-Planck length ${l_{TPl}}$ [52]. It is conceived also that this scale may be used for refinement a basic guideline in many quantum gravity theories [5, 19, 28, 67, 94, 114]. 

However, there is a point which has been often emphasis that lower limit of space-time parameters cannot be performed with precision higher than the Planck scales (see e. g. [63]; and references therein) and believed that \textit {``the Planck scales are limits''} [13]. Nevertheless, it is truly that we are not nearly \textit {``know any basic principle which characterizes the concurrence of General Relativity and Quantum Mechanics, in a similar way as the locality principle does for Special Relativity and Quantum Mechanics''} [34].

To make above our point precise, note that Quantum Cosmology character of time and length parameters of ${t_{TPl}}$ and ${l_{TPl}}$ by the Heisenberg uncertainty principle, losing their applicability only at Universe mass${M_U} > \sim{10^{55}}kg$, ${t_{TPl}} < \sim{10^{ - 106}}s$ and ${l_{TPl}} < {10^{ - 96}}m$, respectively ([52], Sec. 5). 

As a result, at the \textit {TPE} take place \textit {the modified quantum cosmological} processes at which the Universe filled with the $\sim3 \cdot {10^{125}}$gravitons. Moreover, by a \textit {TPl} scenario in the early Universe creation of any one particle with masses exceed ${m_{Gr}}$ cannot be arises. Furthermore, prior to the \textit {``formation of Planck particles''} in the Universe, matter must have existed in the form of string and graviton particles. The fact that, obtained under the assumption (26) creation of string particles appear just at after of $t \sim 2 \times {10^{ - 169}}s$, which take place beyond the Trans-Planck epoch, which we named as Over Trans-Planck epoch (\textit {OTPE}) [52]. Yoneya [151] argued that the generalized uncertainty principle in String theory is not generally valid. In the cited above our work (Sec. 5.4), we also note that \textit {``in the string case, the quantizing gravity by the concepts of the Heisenberg's uncertainty principle losing their applicability''}. However, the string parameters are closely connected to each other Trans-Planck and Planck parameters in a variety of ways, depending on the direct manner in which we compare them. Then, using the relation (26) for the \textit {OTPE} redshift it is argued that 
\begin{equation}
{M_U}/{m_{Str}} \equiv {t_{Str}}/{t_{TPl}} \equiv  \cdots  \equiv {z_{OTPl}} \equiv {N_{Str}} \equiv z_{Pl}^3 \equiv N_{Pl}^3 = {1.4748_{9}}(26) \times {10^{188}},
\end{equation}
where ${N_{Str}}$ is the number of String particles generated in the \textit {OTPE} of Universe. 

In conclusion this paragraf note that the Standard model does not explained neither the mass, nor other parameters of the physical particles [59, 71].

\section{Direct Prediction of the Quantized Dark Energy and Dark Matter }
\label{sec:1}
\subsection{The history of problem}
As discussed in the extensive studies (see e. g. [28, 35, 47, 76, 99, 100, 105, 111, 112, 128]) a large number of different theoretical models exist for Dark Energy and Dark Matter. But up to the present, an entirely convincing theoretical breakthrough has not yet been achieved. In addition, Paul Steinhardt and Neil Turok [125] wrote in 2004: \textit {``Dark energy has shattered that dream. Dark energy was not anticipated and plays no significant role in the theory''}. 

\textit{Observations have forced us to add dark energy ``ad hoc''} [123, 124, 140]. In fact, these first treatements  of course are not accurate. More recently (a ten years later), Turner in discussing this problem point out that ([78]; see also, [138]) \textit {``Dark Energy may be the most profound problem in all of science today''}. Li Miao et al., also at the extreme final point of paper [86] are writing: \textit {``It is without any doubt that the processes of detecting the nature of dark energy and understanding its origin will prove to be one of the most exciting stories in modern science''}, and so on. These historical sayings may be continued. For a recent outlook see also [53].

\label{sec:2}
\subsection{Joint solution of MFRS equations for the three epochs }
Thus, these results show that there are bound to be a radical change in the correct construction of this problem. Nevertheless, in preceding paper [53] as well in a this work we shows that, in an alternative to the Inflationary cosmological model [58, 87, 88] and Standard model of Big bang (see e. g. [15, 47, 91, 105]), problem of origin, nature and quantities DM and DE can be very precisely resolved. With the data from the cited paper we see that, for a more quantitative calculation of the energy distribution in the Universe subparts, it is necessary simultaneous solutions of three above discrete-scaling system of MFRS equations for the present, the Planck's and the Trans-Planck's bounce epochs.

Nowadays, in spite of the abrupt jumps by going from a Trans-Planck's MFRS ${z_{TPl}} = z_{Pl}^2$ to a present structure with ${z_0} = 1$, these equations are linked with the other cosmological parameters via the constant ${\Lambda _0}$ by
\begin{equation}
\frac{{{u_\nu }({t_{TPl}})}}{{{z_{TPl}}({t_{TPl}})}} \equiv \frac{{{u_\nu }({t_{Pl}})}}{{{z_{Pl}}({t_{Pl}})}} \equiv  \cdots  \equiv \frac{{{u_\nu }({t_0})}}{{{z_0}({t_0})}} \equiv {\rho _\nu }({t_0}) \cdot {c^2} \equiv \frac{{{c^4}{\Lambda _0}}}{{8\pi G}} = {6.6038_{8}}(40) \times {10^{ - 14}}J/{m^3},
\end{equation}
where ${z_0}({t_0})$ is defined as ${z_0}({t_0}) = 1$ for the present observable Universe, and ${\Lambda _0}$ is an Present quantum cosmological constant (See also, Table 1).

In addition to Eq (32), the complete state of the homogeneous Universe for various times ${t_i}$, similar to the current value of the energy density parameter ${u_\nu }({t_0})$, can be described also as
\begin{equation}
{u_\nu }({t_i}) = {\rho _\nu }({t_i}) \cdot {c^2} = ({T_i}^4/8\pi ){c^2}/G{\schwa ^2}.
\end{equation}

Numerical evolutions of three epochs MFRS equations by formulas (1), (15) and (20) as a function of cosmic time t, started out from ${t_{TPl}}\sim{10^{ - 106}}s$ to today${t_0}\sim{10^{19}}s$, and temperature started out from ${T_{TPl}}\sim{10^{64}}K$, lead to ${T_0}$ after the Big Bang, in logarithmic scale are plots in Fig.1. (In such a double-logarithmic scale situation, any one power relationships are expressed linearly [65]. In these cases, one expected that the basic structure of our observable Universe in cosmological expansion, is characterized by the three discrete expanding quantized time steps of $\Delta t\sim{10^{ - 62}}$s ,each which characterized by an increase in the Planck redshift scale, with a factor of about $\sim{10^{62}}$. However, the characteristic manifolds of a system are invariant under the transformations of temperature and time, i. e., \textit {products of the square of temperature on time defined by Eqs. (4) are remains constant for any one discrete expanding space-time epoch and phase of the Universe}. 

\label{sec:3}
\subsection{Identification and calculations of Dark Matter and Dark Energy}
In Figure 1 we show the two quantity illustrated combinations of the \textit {``cosmic energy triangle''} in the ${\log _{10}}T - {\log _{10}}t$ plane. The little triangle, for the Planck epoch of space-time, is constructed on the basis of Planck redshift relation (15). Also, a large logarithmic triangle including the very early epoch of space-time is build on the basis of Trans-Planck redshift relation (20). The area of greater logarithmic triangle corresponds to the total energy of the Universe --${M_U} \cdot {c^2}$. 

From these \textit {``cosmic energy triangles''} we can construct some large dimensionless logarithmic ratios. Let us consider the area of little triangle. According to constructing an area of right --angled triangle, the logarithm of the area of little triangle is given by
\begin{equation}
{\log _{10}}S({z_{Pl}}) \equiv {\log _{10}}\frac{{{t_0}}}{{{t_{Pl}}}} \cdot {\log _{10}}\frac{{T_{Pl}^{'}}}{{{T_0}}} \equiv {\log _{10}}{z_{Pl}} \cdot {\log _{10}}z_{Pl}^{1/2} = 1967.0811,
\end{equation}
where $S({z_{Pl}})$ is the component of cosmic energy of the expanding Universe \textit {``enclosed'}' in little triangle and corresponding to the Planck and present epochs position.

\begin{figure}
\centering
\includegraphics[width=1\linewidth]{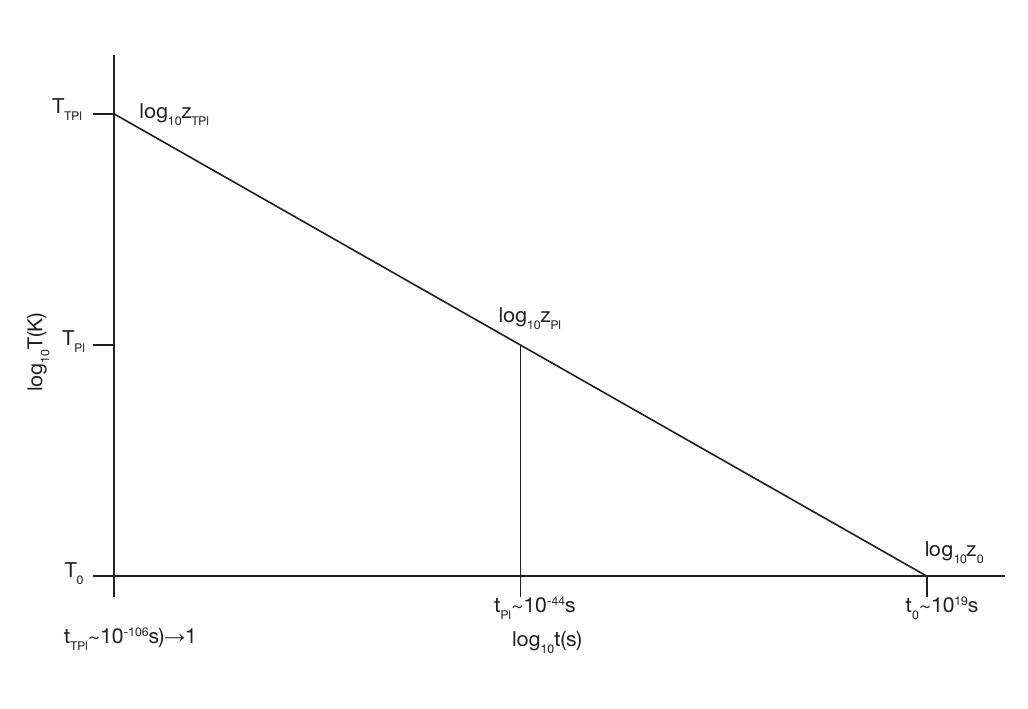}
\caption{The phase diagram include the ${\log _{10}}T(K) - {\log _{10}}t(s)$ dependence of the Planck's MFRS parameters for the steps n = 2, 1, 0 and under the assumption that the current Universe is very close to a spatially flat with ${\Omega _{DM}} + {\Omega _{DE}} \cong {\Omega _{total}} \cong 1$.}
\label{ris:qrafik1_1}
\end{figure}

The same estimation leads to the logarithm of the full cosmic energy of the Universe predicted in a given cosmological model, corresponding to a maximal area of a large triangle by $S({z_{TPl}})$, and determining as
\begin{equation}
{\log _{10}}S({z_{TPl}}) \equiv {\log _{10}}\frac{{{t_0}}}{{{t_{TPl}}}} \cdot {\log _{10}}\frac{{{T_{TPl}}}}{{{T_0}}} \equiv {\log _{10}}z_{Pl}^2 \cdot {\log _{10}}{z_{Pl}} = 7868.3244.
\end{equation}

The resolutions of Eqs.(34)-(35) and that follows is discussed in detail in [53]. We present here the crucial feature.

Dividing a logarithm of the area of little cosmic triangle (34) to a logarithm of the area of a large cosmic triangle (35), the present dark matter energy (DM) density ${\Omega _D}_M$ can be defined by
\begin{eqnarray}
{\Omega _{DM}} \equiv \frac{{{{\log }_{10}}S({z_{PL}})}}{{{{\log }_{10}}S({z_{TPl}})}} \equiv \frac{{{{\log }_{10}}\frac{{{t_0}}}{{{t_{Pl}}}} \cdot {{\log }_{10}}\frac{{T_{Pl}^{'}}}{{{T_0}}}}}{{{{\log }_{10}}\frac{{{t_0}}}{{{t_{TPl}}}} \cdot {{\log }_{10}}\frac{{{T_{TPl}}}}{{{T_0}}}}} \equiv \frac{{{{\log }_{10}}\frac{{{t_0}}}{{{t_{Pl}}}} \cdot {{\log }_{10}}\frac{{T_{Pl}^{'}}}{{{T_0}}}}}{{4{{\log }_{10}}\frac{{{t_0}}}{{{t_{Pl}}}} \cdot {{\log }_{10}}\frac{{T_{Pl}^{'}}}{{{T_0}}}}} \equiv \nonumber \\
\equiv \frac{{{{\log }_{10}}{z_{Pl}} \cdot {{\log }_{10}}z_{Pl}^{1/2}}}{{{{\log }_{10}}z_{Pl}^2 \cdot {{\log }_{10}}{z_{Pl}}}} \equiv \frac{{{{\log }_{10}}z_{Pl}^{1/2}}}{{{{\log }_{10}}z_{Pl}^2}} \equiv  \cdots  = \frac{{1967.0811}}{{7868.3244}} = \frac{1}{4} = 0.25. 
\end{eqnarray} 

The logarithm of cosmic energy \textit {``enclosed''} in the trapezium corresponding to the Trans-Planck and the Planck epochs position of the early Universe, appearing in Fig.1, is given by
\begin{eqnarray}
{\log _{10}}({S_{Trapez}}) \equiv {\log _{10}}S({z_{TPl}}) - {\log _{10}}S({z_{Pl}}) \equiv \nonumber \\
 \equiv {\log _{10}}{z_{Pl}}({\log _{10}}z_{Pl}^2 - {\log _{10}}z_{Pl}^{1/2}) = 5901.2433.
\end{eqnarray} 

From the ratio of the latter to the (35) may eventually inferred a second key -- physical understanding of Dark Energy (DE) density ${\Omega _{DE}}$ defined in the form
\begin{eqnarray}
{\Omega _{DE}} \equiv \frac{{{{\log }_{10}}S(Trapez)}}{{{{\log }_{10}}S({z_{TPl}})}} \equiv \frac{{{{\log }_{10}}{z_{Pl}}({{\log }_{10}}z_{Pl}^2 - {{\log }_{10}}z_{Pl}^{1/2})}}{{{{\log }_{10}}z_{Pl}^2 \cdot {{\log }_{10}}{z_{Pl}}}} \equiv \nonumber \\
\equiv 1 - \frac{{{{\log }_{10}}z_{Pl}^{1/2}}}{{{{\log }_{10}}z_{Pl}^2}} \equiv  \cdots  = \frac{{5901.2433}}{{7868.3244}} = \frac{3}{4} = 0.75.
\end{eqnarray} 

It's worth noticing that the relations in (37) and (38) may also be replaced by the ratios of the time and temperature defined in (34) and (35). Then, we precise estimate an analytic expressions ${\Omega _D}_M$ and ${\Omega _{DE}}$ densities via the \textit {scaling solution parameters} of the three Planck MFRS for the \textit {``quantized''} cosmological epochs. Fig.1 and Eq. (38) shows that, ${\Omega _{DE}}$ is available in the early Universe at times interval ${t_{TPl}} \le \Delta {t_{DE}} <  < {t_{Pl}}$ and length scales${l_{TPl}} \le \Delta {l_{DE}} <  < {l_{Pl}}$. Then, Dark Energy is some initial sort of invisible energy pervading which corresponding prior to Planck epoch of the  Universe at time interval $\Delta {t_{DE}}$ and this \textit {``is bad for Astronomy''} [149]. For the ${\Omega _{DM}}$ case, we have time interval ${t_{Pl}} < \Delta {t_{DM}} \le {t_0}$ and length scales ${l_{Pl}} \le \Delta {l_{DE}} < {l_0}$.These limits corresponds to the MFRS scales ${z_{TPl}} \le z < {z_{Pl}}$ and ${z_{Pl}} < z \le {z_0}$, respectively. This mechanism is contrary to ideas \textit {``in which the Electro-Weak (EW) scale is the large scale, and dark energy scale is the low energy scale''} [73].

\textit {Our results means that Dark energy density (${\Omega _{DE}}$) is appropriate to the invisible early phase of the expanding Universe, including the energy up to the Planck time ${t_{Pl}}$, after the Big Bang. And, correspondingly, Dark Matter density (${\Omega _{DM}}$) is appropriate to the other parts of the Universe energy following a Planck's time ${t_{Pl}}$, up to the present phase of cosmic expansion}. 

Thus, in the model, \textit {where researches ``adopt the point of view that the Planck scales are limits: nothing can go below Planck time and Planck length, and no single particle can go beyond Planck energy''} [13] the search of Dark Energy is unthinkable!

In particular, derived above shape of $S({z_{TPl}})$ and $S({z_{Pl}})$implies that the quantum relativity cosmological parameters ${t_{TPl}}, \cdots ,{t_0}$ and ${T_{TPl}}, \cdots ,{T_0}(CBR)$, corresponding to the testable by observation predictions Dark Matter density ${\Omega _{DM}}$ and Dark Energy density ${\Omega _{DE}}$ of the Universe, also are known quite accuracy. This also \textit {``suggests… that Dark Energy may somehow reflect the unification of  gravity with the other fundamental forces, and hence, paradoxically, physics at energies far above those that can be probed directly with accelerators''} [14, 52, 149].

It is worth noting that the Current Standard $\Lambda$CDM Cosmological model of the Universe \textit {mistakenly presented Dark Energy density ${\Omega _{DE}}$ as a Einstein's Cosmological constant $\Lambda$} ([1, 16, 25, 28, 31, 48, 85, 99, 104, 105, 152] and others), and the Dark Matter density ${\Omega _{DM}}$ as a ${\Omega _{DM}} = {\Omega _{CDM}} + {\Omega _B}$ ($CDM=Cold \ dark \ matter$), with an approximate values of $\Lambda  \approx 0.74$, ${\Omega _{CDM}} \approx 0.21$, and ${\Omega _B} \approx 0.04$ ([25, 77], and references therein). The curvature of a flat Universe for today is given by ${\Omega _k} = 1 - {\Omega _D}_M - {\Omega _{DE}} = 0$. 

Above ``identity'' raises a very deep physical and cosmological problem (see, i.e. [138]). In particular, it requires that the value of Dark Energy density expressed by $0.75$ [53], hereafter denoted by ${\Omega _{DE}}$, and do not $\Lambda $, or ${\Omega _\Lambda }$.

When the best-fit predicted results (36) and (38) are combined they provide that there is the assumption of a flat Universe and that for the present time total mass-energy budget we have a constant value 
\begin{equation}
{\Omega _{DM}} + {\Omega _{DE}} \equiv {\Omega _{tot}} \equiv 1,
\end{equation}

Thereby, we hope that the precise calculation, which corresponds to endowing the Dark Energy and the Dark  Matter densities  given above, which corresponds to endowing the Dark Energy and Dark Matter densities will provide a basis for the concrete formulation on possible construction of our observable Universe in the time interval between ${t_{TPl}}$ and ${t_0}$. \textit {These results are remarkable for many reasons, including its unique fine current observational proof similar to the famous Pythagorean Theorem}, although this theorem is strictly realize only for the extremely small fields [81, 148].

\section{To Understand the Parameter $w$}
Considering $P = {p_\nu }(t)$ as a pressure parameter we must have [109]
\begin{equation}
{u_\nu }(t) =  - {p_\nu }(t) = {c^4}\Lambda /8\pi G.
\end{equation}

Then as a more exotic alternative for the equation of state parameter is [15, 25, 47, 66, 117, 139]
\begin{equation}
w(t) \equiv  - {p_\nu }(t)/{u_\nu }(t) \equiv  - {p_\nu }(t)/{\rho _\Lambda }(t) \cdot {c^2}.
\end{equation}

If accept at face values ${\rho _\Lambda }({t_0})$ and ${u_\nu }({t_0})$ from Table 1, using Eq. (40) for today-vacuum dominated case, we infer from Eqs. (10) and (41) that the Dark energy equation of state parameter has a constant value$w({t_0}) =  - 1$. The independent analyze of a flat Universe confirms that equation of state parameter of the dark energy $w(t_i)$ for a flat cosmology can be (e. g. [23, 131] and references therein) assumed to be constant in time. 

For example, applying the relation (41) to the Trans-Planck epoch we derived
\begin{equation}
w({t_{TPl}}) \equiv  - {p_\nu }({t_{TPl}})/{u_\nu }({t_{TPl}}) \equiv  \cdots  \equiv  - \frac{{{c^4}{\Lambda _{TPl}}}}{{8\pi G}}:\frac{{a_{TPl}^2}}{{8\pi G}} \equiv  - \frac{{{{(c/{t_{TPl}})}^2}}}{{a_{TPl}^2}} =  - {1.00_1} \pm 0.02.
\end{equation}

For other cosmological models of dark energy, $w(t)$ can differ from $-1$ and vary in time. Then, by ${\rho _\Lambda }(t)$ and ${u_\nu }(t)$ parameters from Sec.3, we postulate that parameter $w =  - 1$ is invariant under the Planck epoch.

\section{Comparison with the Observations}
Using data from the [66], we have identified values of our two theoretical components ${\Omega _{DM}} = 0.25$ and ${\Omega _{DE} } = 0.75$ from Eqs. (36) and (38) with the fundamental sets of eleven high-red shifted Supernovae SNe Ia observed with the \textit {Hubble Space Telescope (HST)}, which yield ${\Omega _D}_M = 0.25_{ - 11}^{ + 10}$, and ${\Omega _\Lambda } = 0.75_{ - 11}^{ + 10}$, under the assumptions of a flat universe and that a constant value w=--1. This nicely illustrates that the initial The Supernova Cosmological Project observational data [66] are fully coincides with predictions of theoretical model, given Eqs. (36) and (38), if cosmological constant expected like a Dark Energy. However, in the question that these data \textit{``confirms previous supernova evidence for an accelerating Universe''}, there is a significant disparity on the basis of our MFRS results (see, Sec. 10).

The WMAP3 measurements of the CMB together with the Sloan Digital Sky Survey (SDSS) provides [134], at ${H_0}(km/s/Mpc = 72(3)$ and age of the Universe ${t_0} = 13.8(2)Gyr$ the following accurate values${\Omega _\Lambda } = 0.757(20)$, ${\Omega _{DM}} = 0.243(20)$ and ${\Omega _0} = 1.003(10)$. Our results show good agreement also with the results of [122], which from the combinations of WMAP, SDSS, 2dFGRS, and SNe Ia data finds $w =  - 1.08(12)$. These data do imply that, there is evidence that in the Trans-Planck and Planck epochs of the very early Universe space-time is characterized as the Euclidean (flat) space, coming predominately from the first studies of the CMB measurements (e.g. [11, 122]). 

According to the recent combined analysis of the WMAP 5-yr results [61, 68] the current Universe is consistent with being flat at the 1 \% level, i.e., ${\Omega _{tot}} \cong 1$. Our predicted result ${\Omega _{tot}} \equiv {\Omega _{DM}} + {\Omega _{DE}} = 1$, practically coincides also with a measured value ${\Omega _{tot}} = 0.996_{ - 0.016}^{ + 0.015}$ of the critical flatness density recently derived in the WMAP 7-year data for the total density of normal matter DM, and DE [64, 69, 80]. 

Observationally, the precision measurements of ${\Omega _{DM}}$ and ${\Omega _{DE}}$ are reduced to determine the Hubble constant ${H_0}$ with high accuracy [118]. Different methods of determinations and results of measurements these parameters summarized in [3]. The mean value of Dark Matter and Dark Energy measurements by the WMAP7 Collaborations [69] is
\begin{equation}
{\Omega _{DM}} = 0.27\mbox {, and } {\Omega _{DE}} = 0.71.
\end{equation}

The latter values for the Dark Matter and Dark Energy density based on a new ``the Millienum-II Simulations'' recommended values [18] are 
\begin{equation}
{\Omega _{DM}} = 0.25\mbox {, and } {\Omega _{DE}} = 0.75.
\end{equation}

Thus, it follows that, above theoretical predictions by Eqs. (36) and (38) are in complete agreement with recent results of considerable statistical data.

\section{Age of the Universe }
Substituting for ${t_0}$ value given by Table 1 and from value $1yr = 3.15569259747 \cdot {10^7}s$, for the total age of the present Universe we have 
\begin{equation}
{t_0} = \schwa /T_0^2 = {2.8484_{81}}(16) \times {10^{19}}s = {9.0264_9}(51) \times {10^2}Gyr.
\end{equation}

This value is about of 63.6 times of magnitude older than ``the dynamical age'' $14.2_{ - 0.8}^{ + 1.0}Gyr$ discovered by Riess et al. [115] and 60.6 times of magnitude greater than age $14.9_{ - 1.4}^{ + 1.4}Gyr$ derived by Perlmutter et al. [108] for a flat space-time. The best fit values, the SCP group suggests \begin{equation}
0.8{\Omega _M} - 0.6{\Omega _\Lambda } \cong  - 0.2 \pm 0.1,
\end{equation}
which for a flat model gives
\begin{equation}
{\Omega _M} \cong 0.28\mbox { and } {\Omega _\Lambda } \cong 0.72.
\end{equation}
The best-fit value the HZT for the flat case also is ${\Omega _M} \cong 0.28$ and density parameter ${\Omega _\Lambda } \cong 0.7$.

The Eq. (46) do not change if only our theoretically derived ${\Omega _{DM}}$ and ${\Omega _{DE}}$ results are used. Since our prediction from Sect.4, within the observational errors coincides with the Eq. (46) results
\begin{equation}
0.8{\Omega _{DM}} - 0.6{\Omega _{DE}} \equiv  - {0.2_5}.
\end{equation}

The result derived from the WMAP1 observations under the flat space-time assumptions defined an age of the Universe as ${t_0} = 13.7(2)Gyr$ [11]. The best present (\textit {dynamical}) age for the Universe by WMAP7 is [69]
\begin{equation}
{t_{W7}} = (13.75 \pm 0.13)Gyrs.
\end{equation}
With the above value of $1yr$ we get
\begin{equation}
{t_{W7}} = 13.75(13) \times {10^9}yr = 4.34(41) \times {10^{17}}s,
\end{equation}
from which one can see for the ${t_0}/{t_{W7}}$ ratio
\begin{equation}
{t_0}/{t_{W7}} \equiv {z_{W7}} \equiv  \cdots  = 65.6333(6).
\end{equation}

This means that the age of the Universe by the WMAP observations for SNe Ia objects is evenly small than in the case of the total age of the today Universe${t_0}$. Or, the today Universe ${t_0}$ based on the relation (45) is a factor 65.633 larger than adopted dynamical age of the Universe ${t_{W7}}$ from the (49) and (50). It can be inferred that this predict difference is connected with the Hubble constant problem [15, 56, 69, 118]. In addition, the parameter ${t_0}/{t_{W7}}$ from the ratio (51) within the errors of determination of ${t_0}$ coincides with the adopting value of 100h, commonly employed in the Hubble constant expression of ${H_0} = 100h \cdot km \cdot {s^{ = 1}}Mp{c^{ - 1}}$. With above age value of ${t_0}$ from Eq. (45) and $100h = {t_0}/{t_{W7}} = 65.63(6)$ we determine $h = {0.656_{33}}(6)$ and${H_0} = {65_{63}}(6) \cdot km \cdot {s^{ - 1}} \cdot Mp{c^{ - 1}}$. Then, it turned out that the parameter ${t_0}/{t_{W7}}$bounded with MFRS ${z_{W7}}$ as given in Eq. (51) can be written as
\begin{equation}
{t_0}/{t_{W7}} \equiv {z_{W7}} \equiv  \cdots  \equiv 65.6333 \equiv 100h.
\end{equation}

In Table 2, we compare our estimate with the results of $h$ some authors. As is seen, the errors bars in the key experiment [1] are very small. This mean that our predict result $h = {0.656_3} \pm 0.033$ similar to PLANCK results.
\begin{table}
\caption {The Hubble constant according to different authors.}
\renewcommand{\arraystretch}{1.8}
\begin{tabular}{ll}
\hline
$h$ $s$ & Author\\
\hline
$0.72 \pm 0.08$ & Freedman et al. (2001) [15]\\
$0.68 \pm 0.07$ & Gott et al. (2001) [56]\\
$0.623 \pm 0.063$ & Sandage et al. (2006) [118]\\
$0.704_{ - 0.014}^{ + 0.013}$ & Komatsu, Smith et al. (2011) [69]\\
0.673$ \pm 0.012$ & Ade et al. (2013) [1]\\
${0.656_3} \pm 0.033$ & In this work\\
\hline
\end{tabular}
\end{table}\\
For the latest case the product of ${H_0} \cdot {t_0}$ can be evaluated as
\begin{equation}
{H_0}{t_0} \approx 1,
\end{equation}
i. e., the inverse of the Hubble constant is the total age of the present Universe ${t_0}$.\\
This equality is consistent with an illustration also in Fig.3 [28] at a flat Universe model with of value${\Omega _{DM}} \equiv \Omega _m^{(0)} = 0.25$.

\section{Multiverses Mass and the Number of Different Universes}
From the previously cosmology literature [32, 89], we little know on exist and nature Multiverse. Current best understanding what constitutes the evidence for Multiverse was developed in during the ensuing years [6, 24, 30, 37, 38, 113, 126, 133, 147]. In particular, Ellis [39] noted that \textit {``the multiverse idea is not probable either by observation, or as an implication of well established physics. It may be true but cannot be shown to be true by observation or experiment. Continuation beyond horizon is fine-but just the same old universe! (of horizon on earth). However it does have great explanatory power: it does provide an empirically based rationalization for fine tuning, developing from known physical principles.''}

In Sec.2 we revealed that hypothetical Multiverse masses can be regarded a consequences of predictions of the universal Planck MFRS wrinkles at $n = 2(z_{Pl}^2)$ as ${M_U}$, at $n = 3$($z_{Pl}^3$) as ${M_{PMV}}$, and at $n = 4$($z_{Pl}^4$) as ${M_{BMV}}$, respectively. Actually, by the our model the mass of observable Universe originated at following quantum cosmological temperature and time
\begin{equation}
{M_U} \equiv (\hbar /\schwa  \cdot {c^2})T_{TPl}^2 \equiv (\hbar /{c^2}) \cdot t_{TPl}^{ - 1},
\end{equation}
or as an energy based on the Einstein presentation in the form (see, also [33])
\begin{equation}
{M_U} \cdot {c^2} \equiv  \pm (\hbar /\schwa ) \cdot T_{TPl}^2 \equiv  \pm \hbar /t_{TPl}.
\end{equation}
consistent from the existences of ``Mini Multiverse''(ensemble of universes) with the mass ${M_{PMV}}$, and MFRS's the $z_{Pl}^3 \equiv {z_{OTPl}} \equiv N_{Pl}^3$, ``Big Multiverse'' with the mass ${M_{BMV}}$, and with MFRS $z_{TPl}^2 \equiv z_{Pl}^4 \equiv N_{Pl}^4\sim{10^{250}}$. 

How large is $z \equiv N$ likely to be? Linde and Vanchurin [90] summarized number $\sim{10^{500}}$ as the \textit {``popular estimate of the total number of different universe.''} Using Eq. (19) the derived $z_{TPl}^4$ we can be generated to Planck numbers of \textit {``some ${10^{500}}$ possible vacua of an underlying superstring theory''} [79, 130] by a relation
\begin{equation}
z_{TPl}^4 \equiv N_{TPl}^4 \equiv N_{Pl}^8 \approx {10^{501}}.
\end{equation}

Then, quoted above hypothetical masses of the Mini Multiverse ${M_{PMV}}$ equals to
\begin{equation}
{M_{PMV}} \equiv {z_{Pl}} \cdot {M_U} \equiv {N_{Pl}} \cdot {M_U} \cong {10^{63}}{M_U},
\end{equation}

Thus for ${M_{BMV}}$ related to the Trans-Planck's MFRS by Eq. (19), we have at once
\begin{equation}
{M_{BMV}} \equiv z_{TPl}^2 \cdot {M_U} \equiv N_{Pl}^4 \cdot {M_U}\sim{10^{250}} \cdot {M_U}.
\end{equation}

We can now determine directly \textit {``Mega Multiverse''} with the mass ${M_{MMV}}$ and with Plank's MFRS equation $z_{TPl}^4 \equiv z_{Pl}^8 \equiv N_{Pl}^8 = 6.1 \times {10^{501}}$, giving
\begin{equation}
{M_{MMV}} \equiv z_{TPl}^4 \cdot {M_U} \equiv z_{Pl}^8 \cdot {M_U} \equiv N_{Pl}^8 \cdot {M_U}\sim{10^{502}}{M_U}.
\end{equation}
(As is seen, the mass of ${M_{MMV}}$ may be predicted also by a \textit {``seesaw mechanisms''} [54]).

\label{sec:1}
\subsection{Wrinkles Planck's MFRS equations and physical meaning of the big number}
The formation of our Universe may originate from discrete Planck's MFRS by the follows schematic sketch
\begin{equation}
\cdots (z_{TPl}^4\sim{10^{500}}) \to (z_{TPl}^2\sim{10^{250}}) \to z_{Pl}^3 \to z_{Pl}^2 \to {z_{Pl}} \to z_{Pl}^0({z_0}) \to z_{Pl}^{ - 1} \to z_{Pl}^{ - 2} \cdots .
\end{equation}

We next consider one scenario on the larger number, which have been discussed by Herman Nicolay [98] in the following way: \textit {``To conclude let me restate my main worry. In one form or another, the existing approaches to quantum gravity suffer from a very larger number of ambiguities so far preventing any kind of prediction with which the theory well stand or fall. Even at the risk of sounding polemical, I would put this ambiguity at ${10^{500}}$ (or even more) -- in any case a number too large to cut down''} then, it should be noted that a crucial quantity for phenomenology of larger and infinitely smaller numbers of cut down, possible are bound by the Planck MFRS's $z_{TPl}^4 \equiv N_{TPl}^4\sim{10^{501}}$ and $z_{TPl}^{ - 2} \equiv N_{TPl}^{ - 2}\sim{10^{ - 250}}$ from Equation (15) which are defined as a relevant combinations of $T_{TPl}^2$, ${t_{TPl}}$ and the \textit {quantum cosmological constant} $\schwa $ by relations (4).

\label{sec:2}
\subsection{Conversion of Planck mass in the different epochs}
The conversion of ``cosmological particles'' in these epoch transfers can be treated as a four ``quantized cosmological multiple mass'' transitions from one to the other in follows the order
\begin{equation}
N_{Pl}^2 \cdot {m_{Str}} \to {N_{Pl}} \cdot {m_{Gr}} \to {M_U}/{N_{Pl}} \to {({M_U} \cdot {m_{Gr}})^{1/2}} \to {(\hbar c/G)^{1/2}} \equiv {m_{Pl}}.
\end{equation}

Hence, each of these discrete multiple mass of (61) is identical to the classical ``Planck mass'' of ${m_{Pl}}$, which remains constant in all times [41, 110] of the Universe evolution after the Big Bang. The total numbers of Planck mass (setting${N_{Pl}} \equiv {z_{Pl}}$) in the Universe after $t = {t_{Pl}}$ gets a number somewhere around of${N_{Pl}}\sim5 \times {10^{62}}$. In this case, the Planck redshifts $z_{Pl}^3,z_{Pl}^4, \cdots z_{Pl}^8$ corresponding to the bounce steps of ${10^{62}}e - $folds time intervals between $t\sim{10^{ - 160}}s$ and $t\sim{10^{ - 480}}s$, sandwiched between Big Bang and of $t = 0$ Friedmann model of universe. 

\label{sec:3}
\subsection{Constancy of the total mass (energy) of the Universe in the different epochs}
To summarize, we consider the mass of the Universe in various shape.\\
1. Assuming that Trans-Planck epoch responsible for the dark energy, the Universe mass from Eq. (20) can be estimates as a vacuum energy
\begin{equation}
{M_U} \cdot {c^2} \equiv  \pm [(\hbar /\schwa )T_0^2] \cdot {N_{TPl}} \equiv  \pm ({T_{TPl}}/{T_0}) \cdot {E_{Pl}} \equiv  \pm ({m_{Gr}} \cdot {c^2}) \cdot {N_{TPl}} \equiv  \cdots .
\end{equation}

So that in the Trans-Planck epoch of Universe model is scaled to represent the maximal temperature and here is not necessary for determination length scale [34, 63].\\
2. In the next Planck epoch step, we introduce Planck's parameters to the Universe energy as
\begin{equation}
{M_U} \cdot {c^2} \equiv  \pm [(\hbar /\schwa ){(T_{Pl}^{'})^2}] \cdot {N_{Pl}}.
\end{equation}
3. Of course, the corresponding quantum cosmological constant mass of the currently observable Universe${M_U}$, may be presented also as this is given in Eq. (10).

Here, we initially define a more motivated model for the six cosmological ``particles'' as
\begin{eqnarray}
{M_{PMV}} \cdot {t_{OTPl}} \equiv {M_U} \cdot {t_{TPl}} \equiv {m_{Pl}} \cdot {t_{Pl}} \equiv  \cdots  \equiv \nonumber\\
\equiv {m_{Gr}} \cdot {t_0} \equiv {m_{Str}} \cdot {t_{Str}} \equiv {m_{Last}} \cdot {t_{Last}} \equiv \hbar /{c^2}
\end{eqnarray}
that can be considered as a \textit {modified quantum cosmological mass-time relation}, which is compatible with the Heisenberg's uncertainty principle. 

As is noted in ([52], Sec.5) \textit {``macroscopic and microscopic''} [114] cosmological particles \\
${M_U},{m_{Pl}},{m_{Gr}},{m_{Str}}, \cdots $ may be evolved according to relation
\begin{equation}
{M_U}:{m_{Pl}}:{m_{Gr}}:{m_{Str}}:{m_{Last}} \equiv {z_{Pl}} \equiv {N_{Pl}}.
\end{equation}

These masses for the each substance ${m_i}$ are bounded with the corresponding value of origin temperature ${T_i}$ by the precise relations
\begin{equation}
{m_i}{c^2} \equiv  \pm (\hbar /\schwa ) \cdot T_i^2 \equiv  \pm (\hbar /\schwa ) \cdot {z_i} \cdot T_0^2 \equiv  \pm (\hbar /\schwa ) \cdot {N_i} \cdot T_0^2.
\end{equation}

These new relations between mass (energy) and the temperature, differ essentially from the \textit {classical} expression of the particle energy 
\begin{equation}
{m_i}{c^2} = k \cdot {T_i}(thr),
\end{equation}
presented in up-to-date textbook. Here${T_i}(thr)$is the threshold temperature [52, 145].

\label{sec:4}
\subsection{On some important coincidence with the holographic parameters}
\textit {``Assuming that the holographic principle holds''} [62, 144], we compared some our cosmological parameters derived from the works [52, 53] with the results of Verlinde. These comparisons lead us to following interrelated identity
\begin{eqnarray}
{a_V}/2\pi  = {a_0} = {({\schwa ^2} \cdot {A_S})^{ - 1}},\\
{A_V}/4\pi  = l_0^2 = {[{(\schwa  \cdot c)^2}{A_S}]^2} = {(\schwa  \cdot c/T_0^2)^2} = \Lambda _0^{ - 1},\\
{N_V}/4\pi  = {z_{TPl}} = z_{Pl}^2 = {N_{TPl}} = N_{Pl}^2,\\
{E_V}/2\pi  \cong {M_U} \cdot {c^2},
\end{eqnarray}
where represented in the left hand side parameters are from the Verlinde [144]. Thus, there are two generic features between the holographic scenario and Planck's hierarchy of MFRS, which could be a powerful tool for development in the future of a \textit {``new physics''} theory! In particular, Verlinde [144], named $N_{TPl} = N_{Pl}^2$ as an ``number of bit''.\\
As is seen from Eq. (69) value ${A_V}/4\pi $ is the inverse cosmological constant introduced by   Einstein [52]. Then, Einstein \textit {greater blunder ``constant''} holds in the \textit {``new physical''} theory! Henceforth, if we adopted ${A_V} = 4\pi {({\Lambda _0})^{ - 1}} \equiv 4\pi {({l_0})^2}$=${9.16_2}(11) \times {10^{56}}{m^2}$, then it is straightforward area of the spheres of today Universe's surface! 

\section{MFRS in the latest Planck's epochs corresponding to the low-temperature phase}
In conclusion of Section 3, we shall reiterate the point of clarity on understanding of the classical small value $\Lambda  \cdot (G\hbar /{c^3})\sim{10^{ - 123}}$ (see, i.e., [99]). In this light, the last equation of subsection 3.3 established the fundamental cosmological meaning of this famous relation as an inverse Trans-Planck epochs ${z_{TPl}} \equiv z_{Pl}^2$ predicted from basic MFRS model by Eq. (20).

Over many years, there was no clear theoretical picture of what to expect. In 1992 at this George Darwin Lecture J. Barrow stated that \textit {``…It is worth remarking that in the 1930s the largeness of reciprocal of equation $\Lambda  \cdot l_{pl}^2 \le {10^{ - 121}}$ was regarded as a major mystery''} and \textit {``problem by Eddington and Dirac''} [7-9]. So, in accordance with our previous estimates [52], the proposed Planck hierarchy of MFRS model can resolve this very olden and profound cosmological problem. 

Thus, it turned out that the problem which at once directly thinks of Eddington and Dirac at cyclic Universe scenario may be based on ${(z_{Pl}^{ - 2})^{ - 1}} \to z_{Pl}^2$ [in ``classical'' form ${({\Lambda _0} \cdot l_{Pl}^2)^{ - 1}} \to ({\Lambda _0} \cdot l_{Pl}^2)$].

On the other hand, the analysis of Eq. (30) immediately shows us that the above smaller values can be related to the post-Big-Bang steps of MFRS equations in the following correct way
\begin{eqnarray}
z_{TPl}^{ - 1} \equiv z_{Pl}^{ - 2} \equiv N_{TPl}^{ - 1} \equiv N_{Pl}^{ - 2} \equiv {\Lambda _0} \cdot l_{Pl}^2 \equiv (G\hbar /{c^3}) \cdot {\Lambda _0} \equiv  \cdots  \equiv \nonumber \\
 \equiv {({m_{Gr}}/{m_{Pl}})^2} \equiv  \cdots  \equiv {({t_{Pl}}/{t_0})^2} \equiv  \cdots  = {3.5811_{3}}(40) \times {10^{ - 126}}.
\end{eqnarray} 

Note also, that this representation allows us to write some string parameters in terms of purely Planck parameters as
\begin{eqnarray}
z_{Pl}^{ - 2} \equiv {(\frac{{{T_{Str}}}}{{T_{Pl}^{'}}})^2} \equiv (\frac{{{t_{Pl}}}}{{{t_{Str}}}}) \equiv (\frac{{{l_{Pl}}}}{{{l_{Str}}}}) \equiv (\frac{{{m_{Str}}}}{{{m_{Pl}}}}) \equiv {(\frac{{{m_{Gr}}}}{{{m_{Pl}}}})^2} \equiv (\frac{{{m_{Last}}}}{{{m_{Gr}}}}) \equiv \nonumber\\
\equiv  \cdots  \equiv (\frac{{{a_{Str}}}}{{{a_{Pl}}}}) \equiv {(\frac{{{a_0}}}{{{a_{Pl}}}})^2} \equiv  \cdots  = {3.5811_3}(40) \times {10^{ - 126}}.
\end{eqnarray}
Here 
\begin{eqnarray}
{T_{Str}} =  \pm (T_0^2/T_{Pl}^{'}) =  \pm (z_{Pl}^{ - 1} \times T_{Pl}^{'}) =  \cdots  =  \pm {1.2032_7}(13) \times {10^{ - 31}}K,\\
{t_{Str}} = t_0^2/{t_{Pl}} = {1.5050_5}(16) \times {10^{82}}s,\\
{l_{Str}} = l_0^2/{l_{Pl}} = {4.5120_4}(48) \times {10^{90}}m, \\
{m_{Str}} = m_{Gr}^2/{m_{Pl}} = {7.7943_6}(83) \times {10^{ - 134}}kg = {4.37_4}(47) \times {10^{ - 98}}eV/{c^2},\\
{a_{Str}} = a_0^2/{a_{Pl}} = {1.9916_7}(21) \times {10^{ - 74}}m/{s^2},\\
{m_{Last}} \equiv z_{Pl}^{ - 2} \cdot {m_{Gr}} = {1.4741_0} \times {10^{ - 196}}kg = {8.27_4} \times {10^{ - 161}}eV/{c^2}.
\end{eqnarray}

Here ${t_{Str}}$ corresponds to an age of Universe since the Big Bang comprising ``Inverse Trans-Planck Rip''${t_{Str}} = 4.8 \times {10^{65}}Gyr$. So, this is compelling moment in \textit {``the destiny of the Universe''} [139], and is the oldest age limit of our Universe! 

Pre-existing step in this direction by on MFRS model is$z_{Pl}^{ - 1}$, achieved ranging a temperature $\sim6.4 \times {10^{ - 13}}K$ and time $\sim 2 \times {10^{28}}Gyr$. This result obtained based on MFRS model, is $\sim{10^{27}}$ large than \textit {``limit on the minimum time to a (speculative) Big Rip''} $\sim 30Gyr$, derived by Riess et al. [116]. In addition, one of the first estimates of minimal remaining time to the Big Rip, found in [23] is $22 Gyr$. 

Noting that above results, from MFRS with $z_{Pl}^{ - 2}$ is also a $\sim4 \times {10^5}$ times large than predicted for the Big Rip moment $\sim{10^{60}}Gyr$, taken from WMAP7 data [49]. 

In this limiting case, after a break down of all structures and particles [99], in accordance with the \textit {``quantum gravitational uncertainty principle''} [102] the Universe is filled only string particles with the masses ${m_{Str}}\sim7.8 \times {10^{ - 134}}kg$ and numbers of 
\begin{equation}
{N_{Str}} \equiv N_{Pl}^3 \equiv z_{Pl}^3 \equiv {M_U}/{m_{Str}}\sim7 \times {10^{189}}.
\end{equation}
In work [17], with a few exceptions, for an ``absolute minimal mass'' adopted ${M_{\min }} \approx 1.4 \times {10^{ - 124}}kg$. 

Then, it would be possible that in an instant at an expansion of the Universe, the quantum-cosmological fluctuations of the vacuum [70, 136] would reverse the \textit {arrow of time} [106] and the huge Universe can suffer of quantum bounce with the $(z_{Pl}^{ - 2}) \to z_{Pl}^{n > 0}$. This shape is meant that the going from $z_{Pl}^{ - 2}$ to a $z_{Pl}^n$at $n \ge 0$ may be compared with the application of Weyl scenarios $WEYL = 0$ and $WEYL \to \infty $ for quantum fluctuations [107].

On the other hands, there is evidence that particle with the minimal energy ${E_{Min}}$ for both MFRS events are, ${m_{Last}} \cdot {c^2} = ({m_{Gr}} \cdot {c^2})z_{Pl}^{ - 2}$ and the strings with the masses ${m_{Str}} >  > {m_{Last}}$. Eventually, following the break down and the beginning of the Universe may retain of ${N_{Str}}\sim7 \cdot {10^{189}}$ number of string numbers. Nevertheless, the temperatures ${T_{Str}}$ and ${T_{Last}}$ can be changed from a positive to a negative absolute temperature by the above commutation relation (56) in the forms
\begin{equation}
{T_{Str}} =  \pm z_{Pl}^{ - 1/2}{[({m_{Gr}} \cdot {c^2})(\schwa /\hbar )]^{1/2}} =  \pm {[({m_{Str}} \cdot {c^2})(\schwa /\hbar )]^{1/2}} =  \pm {1.20_3}(13) \times {10^{ - 31}}K.
\end{equation}
and
\begin{equation}
{T_{Last}} =  \pm [({m_{Last}} \cdot {c^2})(\schwa /\hbar )] \equiv  \pm z_{Pl}^{ - 1}{[({m_{Gr}} \cdot {c^2})(\schwa /\hbar )]^{1/2}} =  \pm 1.45 \times {10^{ - 62}}K.
\end{equation}

The idea for the rest energy of String particles $ \pm {m_{Str}} \cdot {c^2}\sim4 \times {10^{ - 98}}eV$ supported also by an explanation of the Einstein formula of$E =  \pm m \cdot {c^2}$, which was discussed by Dirac in 1975 [78]. 

As a result, by quantum gravity cosmology conception, in the end of \textit {cyclic and oscillatory} life, the Universe not attained absolute zero temperature as a consequence of the condition that energy in interval $ - m \cdot {c^2} < E < m \cdot {c^2}$ must be barred from use. Furthermore, the fundamental third law of thermodynamics (Nernst's Law) is satisfied and ruled out of singularity of Friedmann model (see, Eq.(3) and schematically of Fig.2). 

Later the universe much like to an our observable Universe is beginning of newly cycle expansion in a set of the big Planck Multiverse with the very massive ``particle''$\sim{10^{62}}{M_U}$, in form of one collapsing with positive and possible symmetric negative energy
\begin{equation}
{M_U} \cdot {c^2} \equiv  \pm ({c^4}/G){l_{Str}} \cdot z_{Pl}^{ - 1} \equiv  \pm (\hbar /\schwa ) \cdot {N_{Str}} \cdot T_{Str}^2 \equiv  \cdots  \equiv  \pm (\hbar /\schwa ) \cdot {N_{GR}} \cdot T_0^2 \equiv  \cdots  \equiv  \pm (\hbar /\schwa )T_{TPl}^2.
\end{equation}

On the other hand, this assumption \textit {``leads us directly to the idea that..., the overall time scale leads us to the conclusion that the universe is cyclic''} [22, 125, 140] and \textit {``in this oscillatory…, our universe will be destroyed and thebe rebuilt again and again''} [148]. In the popular form similar process well described by Joseph Silk in the classical book of \textit {``The Big Bang''} ([120]; see, also [29, 146]). 

Thus, consistent with the three important equations (1), (15) and (20) durations, during the transformation MFRS between $z_{Pl}^2$ and$z_{Pl}^{ - 2}$, makes possible, rewrite equation (4) for the constant of $\schwa $ as 
\begin{equation}
\schwa  \equiv T_{TPl}^2 \cdot {t_{TPl}} \equiv (T_{Pl}^{'})^2 \cdot {t_{Pl}} \equiv  \cdots  \equiv T_0^2 \cdot {t_0} \equiv  \cdots  \equiv T_{BR}^2 \cdot {t_{BR}} \equiv  \cdots  \equiv b({c_2}/{c_1}) \cdot {E_{Pl}}.
\end{equation}

This equation explicitly shows that for all feasible cosmological epochs a set of discrete points Planck's MFRS should have five quantum steps of description by an expansion factor of approximately $5 \times {10^{62}}$, permissible for our Universe by the Heisenberg's uncertainty principle, represented by
\begin{equation}
{Z_{MFRS}} \to {z_{TPl}} \equiv z_{Pl}^2,{z_{Pl}},z_{Pl}^0 \equiv {z_0},z_{Pl}^{ - 1},z_{Pl}^{ - 2}.
\end{equation}

As is shown above the first three bound of Planck MFRS with $n = 2,1$ and 0 from Eq. (85) are suitable for Eq. (83). But in which the product $T_{BR}^2 \cdot {t_{BR}}$ should be replaced by a term $T_{Str}^2 \cdot {t_{Str}}$, or in this case ${z_{BR}} = z_{TPl}^{ - 2} = z_{Str}^{ - 2}$ has no evidence.

The results of this part of our analysis shows that the decomposing of the predicting Universe comes to $z_{Pl}^{ - 2}\sim{10^{ - 126}}$ and age of ~${10^{65}}Gyr$ provided that, ${T_{Str}} =  \pm (\schwa /{t_{Str}})^{1/2}$ is to be near absolute zero temperature, and also by Eqs. (75) and (81), equal to $ \pm 1.2 \times {10^{ - 31}}K$. In addition, it is clear from Eq. (82) that temperature ${T_{Last}}\sim \pm {10^{ - 62}}K$ by the Eq. (84) will approximately corresponds to fantastically finite time of our Universe of ${t_{Last}}\sim{10^{128}}Gyr$.

Eventually, main argument supporting above calculations is the assumption of the Big Bang the evidence of creation $+{M_U} \cdot {c^2}$/$ - {M_U} \cdot {c^2}$ from the Multiverse annihilation back into new Universe.

\section{On accelerated expansion problem}
As is noted above in recent years, once a cosmic acceleration is essentially has been established in observations by two above quoted group's data on SNe Ia [108, 115]. At different time the visible Universe is defined as an accelerated by many researchers [4, 42, 43, 83, 84, 116]. In a similar manner, referring back to SNe Ia data, even if \textit {``the first supernovae results did not yet show acceleration''} [137]. 

Yet, our understanding is that this phenomenon may be physically related directly to speed of light in vacuum $c = 299792458m/s$ [95]. Recall also that the Universe acceleration is the time-dependent value [52]. Then this is phenomena which should be also subject to the physical deciphering. Here we explore whether this exceptional condition can assist in providing an accelerating model. 

Below we shall use the directly measured in [116] and by Seven-Year WMAP (WMAP7) [69] observational data and borrow their results. (Here for the brevity WMAP7 we replaced by ``W7''). \\
However, existence of acceleration must be taken as probable, but not conclusively proved. In the light of the new methods [52] this phenomenon may no longer be valid. 

It is well known that none of the WMAP and SNe Ia measurements in the expansion history of the Universe for the t$ \le {t_0}$ does not give a correct value of a(t) in physical contents (i. e., in $m/{s^2}$ units).Therefore, the total set of physical parameters our consistent system not represents ${a_W}$. Nevertheless, it is conceived that the commoving acceleration ${a_W}$ and dynamical time ${t_W}$ in the observations of SNe Ia are related to the identical predicted background acceleration ${a_0}$ and total age of the present Universe ${t_0}$ by the general definition -- \textit {with the new law of cosmological expansion}, derived earlier ([52], Eq. (5.3)) 
\begin{equation}
{a_{TPl}} \cdot {t_{TPl}} \equiv {a_{Pl}} \cdot {t_{Pl}} \equiv  \cdots  \equiv {a_e} \cdot {t_e} \equiv  \cdots {g_U} \cdot {t_U} \equiv  \cdots  \equiv {a_{W7}} \cdot {t_{W7}} \equiv  \cdots  \equiv {a_0} \cdot {t_0} \equiv  \cdots  \equiv c.
\end{equation}

These equalities suggest that \textit {the velocity of expansion of the Universe represented as ${a_i} \cdot {t_i}$ does not vary with the cosmological time, and is equal to speed of light in vacuum c.} (Here all times and accelerations are reckoned from Big Bang).

\begin{figure}
\center {\includegraphics[width=1\linewidth]{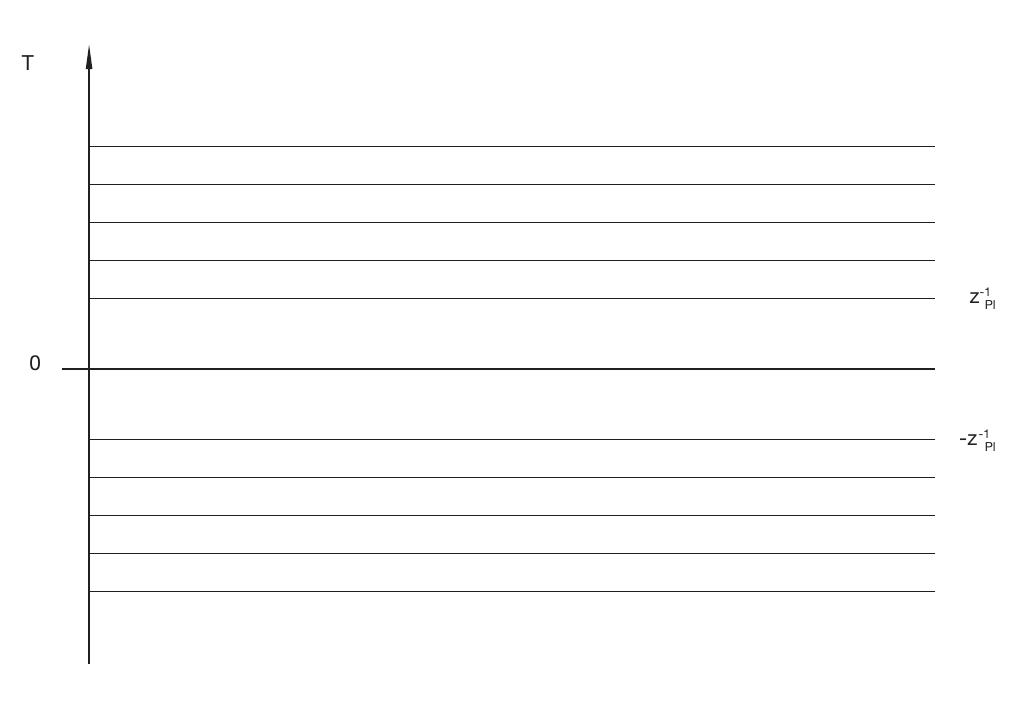}}
\caption{The schematically relation between the temperature and Planck MFRS's. Here ``0'' is the zero absolute temperature state. $z_{Pl}^{ - 1}$ and $ - z_{Pl}^{ - 1}$ are the Planck redshifts which are approach to zero. (This representations of temperature is replete the Dirac graphical diagram for the inertial mass)[33].}
\label{ris:qrafik2_2}
\end{figure}

More specific case, for the other presentation of the early Universe, Eq. (86) can be expected as
\begin{equation}
\cdots  \equiv {a_{Pl}} \cdot (\frac{{{t_{Pl}}}}{{1yr}}) \equiv  \cdots  \equiv {a_{W7}} \cdot (\frac{{{t_{W7}}}}{{1yr}}) \equiv {a_0} \cdot (\frac{{{t_0}}}{{1yr}}) \equiv  \cdots  \equiv \frac{c}{{1yr}} \equiv {g_U} \equiv {9.50005264_{265}}(m/{s^2}).
\end{equation}
 \\

On the other hand, utilizing these equations we deduced that \textit {in all expansion epochs of the Universe the product of cosmological acceleration on this age in this moment, remains constant and is equal\footnote {It is impressive that the value of ${g_U}$ from Eqs. (87)-(88) coincides with acceleration of gravity at the Earth's surface in altitude 9286 m, equivalent to a top of Maunt Everest (8850 m), plus 436 m from sea level, which makes the ${g_U}$ testable. The detailed analysis of a given ``geophysical constant'' for the acceleration of the Earth's surface realm could allow us to establish prediction observation for the all times of evolution of the Universe. Then, apart from the constant $c$ the cosmology should be allowed no less a key observable cosmological-geophysical constant${g_U}$. It seems plausible that existing of ${g_U}$ can be tested with the high-precision accelerometers in terrestrial experiments.} to }

\begin{equation}
{g_U} \equiv {a_i}({t_i}) \cdot ({t_i}/1yr) \equiv \frac{c}{{1yr}} \equiv {9.50005264_{265}}(exact)(m/{s^2}).
\end{equation}

Utilizing above equations for the${t_U} = 1yr$, we can determine other parameters of MFRS, giving 
\begin{eqnarray}
{z_U} = {9.02_5}(50) \times {10^{11}},\\
{T_U} = 2.63(15) \times {10^6}K,\\
{l_U} = 9.46(52) \times {10^{15}}m.
\end{eqnarray}

According to Eq. (86) ratio of accelerations two moments of the Universe is equal to ratio their instant times. Then we have
\begin{equation}
{a_{W7}}/{a_0} \equiv {t_{W7}}/{t_0} \equiv  \cdots .
\end{equation}

With the new laws of expansion we must revise our estimate of the acceleration by using the WMAP7 result from (86) combined with SNe Ia data (see, e. g. [4]). With the value ${t_{W7}}$ from Eq. (92) we get
\begin{equation}
{a_{W7}} = \frac{{{a_0} \cdot {t_0}}}{{{t_{W7}}}} = \frac{c}{{{t_{W7}}}} = 6.91(65) \times {10^{ - 10}}m/{s^2}.
\end{equation}

The existence of such equations between cosmological accelerations and the ``dynamical ages'' would place a fundamental role in quantum cosmology.

In particular, the values of Planck's epoch parameters are the simplest route to probing the acceleration history. In the Planck epoch, ${t_{Pl}} = {(\hbar G/{c^5})^{1/2}}$ is ``the Planck age'' of the Universe. The value ${a_{Pl}} = {c^2}{({c^3}/\hbar G)^{1/2}}$ [52] is the Planck's acceleration for this epoch. Then, for the corresponding Planck epoch by (87) we have 
\begin{equation}
{a_{Pl}} \cdot {t_{Pl}} \equiv c.
\end{equation}

In turn, at ${t_e} = 1.2880886570(18) \times {10^{ - 21}}s$ [95, 96] and ${a_e} = 2.32742099(3) \times {10^{29}}m/{s^2}$ [52], for the electromagnetic time phase, we anew have
\begin{equation}
{a_e} \cdot {t_e} = 299792457.73m/s = c!
\end{equation}

Thus, it is fortunate that the product of the time to acceleration by the identity relations (72) is applicable for the all epochs of the Universe history [52] and is unshakable law. In this way, there are firm physical grounds for assuming that 
\begin{equation}
{a_{TPl}}({t_{TPl}}) >  > {a_{Pl}}({t_{Pl}}) >  >  \cdots  >  > {g_U}({t_U}) >  > {a_{W7}}({t_{W7}}) >  \cdots  > {a_0}({t_0}).
\end{equation}

This means that the cosmological acceleration in each later epoch of evolution of our Universe is less than in proceeding then the much. The fundamental reasons for the origin of ``acceleration'' are assumptions on higher values for Hubble constant and younger age of the Universe defined from the product${H_0}{t_0}$. 

Final result of HST collaboration, ranging over 15 yr based on 62 SN Ia with $3000km/s < zc = {v_{CMB}} < 20000km/s$ and on 10 luminosity-calibrated SN Ia is [118] ${H_0}(\cos mic) = 62.3 \pm 1.3(random) \pm 5.0(systematic)$. Nowadays, however, there are strong opinions that this value can be overestimated [132]. Therefore one can conclude that observational data required \textit {``reliable astrophysical estimates of${H_0}$''} [15]. For example, at value for Hubble constant ${H_i} = 37.4 kms^{ - 1}Mp{c^{ - 1}}$ and age of the Universe ${t_i} = 17.4Gly = 5.49 \times {10^{18}}s$, the ``acceleration'' of the Universe disappears [97]. 

The relations (86), and (95) are fundamental for the reconciliation of the parameters ${t_i}$ and the ${a_i}$. Therefore, the causes of the acceleration of the Universe at the epoch ${t_{W7}} < {t_0}$ would a large ${a_{W7}} > {a_0}$. Thus a cosmic acceleration ${a_{W7}}$, established in early observations [108, 115] and calculated here in Eq.(94), \textit {really appertain to the Universe instant time of $\sim{t_{W7}}$, and not a present epoch time (age)} ${t_0}$, at which the present day acceleration ${a_0}$ (see Tabl.1) is equal to 
\begin{equation}
{a_0} = c/{t_0} = (c/\schwa )T_0^2 = {1.05246_{42}}(61) \times {10^{ - 11}}m/{s^2}.
\end{equation}

Finally, we may argue that in the history of the expanding Universe based on Eqs. (86)-(97), throughout its \textit {all epochs, acceleration as a function of time decreased with increasing of time}.

\section{Recommended values of the Dark Energy and Dark Matter in physical units}
For most cases of interest involving the exact quantities of the dimensionless parameters ${\Omega _{DE}}$ and ${\Omega _{DM}}$. Then, derived parameters along with the constant $\schwa $ from Eqs.(3), (4) and (32) to the list of FPC can be considered as the ``Recommended values of the Fundamental Constants for Cosmology and Astrophysics'' (RFCfCA), which are nonetheless important in the immediate cosmological analysis and applied physics [134, 141]. In this case, these parameters of new cosmological physics can be expressed in SI system or atomic energy units.

Then, our one of other new physical findings it would also seem that the dominant component of the Universe -- the ${\Omega _{DE}}$ (in the physical units${M_{DE}}$) is defined to
\begin{equation}
{M_{DE}} = {\Omega _{DE}} \times {M_U} \cdot  = 8.6250_4(52) \times {10^{54}}kg = 4.8383_0(28) \times {10^{81}}GeV/{c^2},
\end{equation}
where ${M_U}$ is the total mass of the observable Universe given by Eq. (10). The same value for DM is
\begin{equation}
{M_{DM}} = {\Omega _{DM}} \times {M_U} = 2.8750_1(18) \times {10^{54}}kg = 1.6127_7(9) \times {10^{81}}GeV/{c^2}.
\end{equation}

\textit {Eventually ${M_{DE}} \cdot {c^2}$ and ${M_D}_M \cdot {c^2}$ are the Trans-Planck and the Post-Planck epochs energy of our current Universe, respectively}. It is evident that
\begin{equation}
{M_{DE}} \cdot {c^2} = 3{M_{DM}} \cdot {c^2} = 0,75(h/\schwa ) \cdot T_{TPl}^2.
\end{equation}

\textit {Thereby, ${M_{DE}}$ is impressive cosmological parameter, content of the explosive temperature with dimensions of energy}.

In conclusion, above equations contain a four good determined quantum -- ``cosmological particles'' mass -- the ${m_{Str}},{m_{Gr}},{m_{Pl}}$ and ${M_U}$, which at all time related to each other by relation
\begin{equation}
{m_{Str}} = {m_{Pl}} \cdot \frac{{{m_{Gr}}}}{{{M_U}}} \equiv {m_{Gr}} \cdot \frac{{{m_{Pl}}}}{{{M_U}}} \equiv \frac{{{m_{Gr}}}}{{{z_{Pl}}}} \equiv \frac{{{m_{Pl}}}}{{z_{Pl}^2}} \equiv  \cdots .
\end{equation}

Note that expression (101) via graviton mass ${m_{Gr}}$ is meaningful for some physical particles. Employing the definition (G2010)
\begin{equation}
{m_{Gr}} = {m_e}/{z_e} \equiv {m_{{Z^0}}}/{z^0},
\end{equation}
where ${m_e}$, ${m_{{Z^0}}}$are the electron and ${Z^0}$ boson rest masses, and ${z_e}$ and ${z_z}$ are MFRS for the electron and ${Z^0}$ bosons ``epochs'', respectively. For the sake of simplicity, in the case electron, from the combination of (101) and (102) the ${m_{Str}}$ can be determined as 
\begin{equation}
{m_{Str}} = \frac{{{m_{Pl}}}}{{{M_U}}} \cdot \frac{{{m_e}}}{{{z_e}}} \equiv \frac{{{m_{Gr}}}}{{{z_{Pl}}}} \equiv \cdots \approx 7.8 \times 10^{-134} kg.
\end{equation}

These results can play an important role in understanding the future development of the fundamental quantum theory of testable Universe (see, e.g. [60, 114]). (Here will not be considered other relation with the ${m_{Last}}$, which is determined in (61). On the other hand, it is argued that, particle with the ${m_{Last}}$ can be regarded as the first cosmological mass in a creation of matter from the radiation before the Over-Trans-Planck (OTP) epoch).

\section{Concluding remarks }
Finally, as noted in the preceding works [50-53], we with assiduity determined our quantum cosmological parameters better than with $ \pm 10$percent precisions. This type of determinations for the physical measurements, point out William Thompson early as 1900 ([55], see also, [60]).

What our predicted results, in agreement with WMAP observations [11, 61, 68, 69, 80, 84, 122] suggest that, in general our predictions on the total mass-energy of the Universe no significant distinguished from the standard $\Lambda $CDM flat model, with $0.99 < {\Omega _{tot}} < 1.01$ (95\% CL). Then, it appears that via scaling parameters of the three discrete cosmological epochs MFRS, we firstly obtain (see, also [53]) more precise and consistent expressions for the alternative DM and DE densities of the Universe, which one gets to ${\Omega _D}_M = 0.25$, and ${\Omega _{DE}} = 0.75$. Then, the greatest mystery nature of the dark matter, dark energy and the cosmic acceleration (see, e. g. [12, 26, 138]) here is answered.

Thus, we have compelling theoretical evidence for explanations nature the Dark Energy and Dark Matter content of the observable Universe. At the same time, our scenario fit presented in Sect.7 do not indicate any motivation for cosmic acceleration in the Universe history of the expansion [53]. This is a tight connection between quantum cosmology conception and simultaneous scaling solutions of MFRS equations and we have concentrated our efforts on this direction.

\section{Acknowledgments }
This work was made in the Shemakha Astrophysical Observatory, National Academy of Sciences of the Azerbaijan Republic. The authors thank Dr. M. S. Chubey for careful reading and editing of this manuscript. We dedicate this paper to the memory of our dear Dina Gasanalizade, whom we suddenly miss.

\bibliographystyle{spmpsci}      
\bibliographystyle{spphys}       

\begin{thebibliography}{}
%
%
\bibitem{RefJ}
Ade P., Aghanim N., Armitage-Caplan C. et al., ``Planck 2013 results. XVI. Cosmological parameters'', [arXiv:astro-ph. CO/1303.5076v1], (2013)
\bibitem{RefJ}
Aguirre A. and Tegmark M., Multiple universes, cosmic coincides, and other dark matters, [arXiv:hep-th/0409072v2], (2004) 
\bibitem{RefJ}
Albrecht A., Bernstein D., Cahn R. et al, Report from the Dark Energy Task Force, [arXiv:astro- ph/0609591], (2006 )
\bibitem{RefJ}
Amanullah R. et al., Spectra and Hubble Space Telescope light curves of six type Ia supernovae at $0.511< z <1.12$ and the Union 2 compilation, ApJ 716 712, (2010) 
\bibitem{RefJ}
Ashteker A., Pawlowski T. and Singh P., Quantum nature of the big bang: Improved dynamics, Phys. Rev. D 74, 084003 [arXiv:gr-qc/0607039], (2006)
\bibitem{RefJ}
Barrau A., Physics in the multiverse: an introductory review, CERN courers 47, 13 [arXiv:astro-ph/1711.4460v2], (2007)
\bibitem{RefJ}
Barrow J., Unprincipled Cosmology, Q. J. R. Astr. Soc., 34, 117, (1993)
\bibitem{RefJ}
Barrow J. and Tipler F., The Anthropic Cosmological Principle, UOP, Oxford, (1984)
\bibitem{RefJ}
Barrow J., In Modern Cosmology in Retrospect, Chapter 5; eds. Bertotti R., Balbinot R., Bergia S., and Messina A., CUP, Cambridge, (1990)
\bibitem{RefJ}
Bastero-Gil M. and Mersini L., Supernova type Ia data and the cosmic microwave background of modified curvature at short and large distances, Phys. Rev. D 65, 023502, (2002)
\bibitem{RefJ}
Bennett C., Halpern M., Hinshaw G. et al., ``First Year Wilkinson Microwave Anisotropy Probe (WMAP)  Observations: Preliminary Maps and Basic Results'' [WMAP Collaboration], ApJS, 148, 1 [arXiv:astro-ph/0302207v3], (2003)
\bibitem{RefJ}
Bianchi E. and Rovelli C., Why all these prejudices against a constant? [arXiv:astro-ph. CO/ 1002.3966v3], (2010)
\bibitem{RefJ}
Biermann P. and Harms B., Can dark energy be gravitational waves? [arXiv:astro-ph. CO/1305.0498v1], (2013)
\bibitem{RefJ}
Bjorken J., The Classification of Universes Phys. Rev, D 67, 043508, [arXiv:astro-ph/0404233], (2003)
\bibitem{RefJ}
Blanchard A., Evidence for the Fifth Element--Astrophysical status of Dark Energy, Astron. Astrophys. Rev. 18, 595 [arXiv:astro-ph/1005.3765v2], (2010)
\bibitem{RefJ}
Blanchet L. and Le Tiec A., Dipolar Dark Matter and Dark Energy, [arXiv:astro-ph.CO/0901.3114v2], (2009)
\bibitem{RefJ}
Bohmer C. and Harko T., Physics of dark energy, Found. Phys. 38, 216 [arXiv:gr-qc/0602081v6], (2006)
\bibitem{RefJ}
Boylan-Kolchin M., Springel V., White S., Jenkins A. and Lemson G., Resolving Cosmic Structure Formation with the Millienium-II Simulations, Mon. Not. Roy. Astron. Soc., 398, 1150 [arXiv:astro-ph. CO/ 0903.3041v2], (2009)
\bibitem{RefJ}
Brandenberger R. and Zhang X., The Trans-Planckian Problem for Inflationary Cosmology Revisited, [arXiv:hep-th/0903.2065], (2009)
\bibitem{RefJ}
Branderberger R. and Martin J., Trans-Planckian Issues for Inflationary Cosmology [arXiv:astro- ph/ 211.6753v1], (2012)
\bibitem{RefJ}
Browne P., The Case for an Exponential Red Shift Law, Nature, No. 4820, 1019, (1962)
\bibitem{RefJ}
Burbidge G., A Realistic Cosmological Model Based on Observations and Some Theory Developed over the Last Years, [arXiv:astro-ph/0811.2402v1], (2008)
\bibitem{RefJ}
Caldwell R., Kamionkowski M. and Weinberg N., Phantom Energy and Doomsday Phys. Rev. Lett. 91, 71301 [arXiv:astro-ph/0302506v1], (2003)
\bibitem{RefJ}
Carr B. (ed), Universe or Multiverse? CUP, Cambridge, (2007)
\bibitem{RefJ}
Carroll S., The Cosmological Constant, Living Rev. Relativity 4, 1 [arXiv:astro-ph/0004075v2], (2001)
\bibitem{RefJ}
Carroll S., Why is the Universe Accelerating? Carnegie Observatories Astrophys. Ser. 2; Measuring and Modeling the Universe, ed. W. L. Freedman (Cambridge: Cambridge Univ. Press) [arXiv:astro-ph/0310342v2], (2003)
\bibitem{RefJ}
Casado J., Connecting Quantum and Cosmic Scales by a Decreasing Light Speed Model [arXiv:astro-ph/0404130], (2004)
\bibitem{RefJ}
Copeland E., Sami M. and Tsujikawa S., Dynamics of Dark Energy, Int. J. Mod. Phys. D 15, 1753 (2006) [arXiv:hep-th/0603057v3], (2006)
\bibitem{RefJ}
Davies P., SUPERFORCE. The Search for a Grand Unified Theory of Nature. A Touchstone Book. Published by Simon and Schuster, Inc. New York, (1985)
\bibitem{RefJ}
Davies P., Multiverse Cosmological Models Mod. Phys. Lett. A19, 727--744 [arXiv:astro-ph/0403047], (2004)
\bibitem{RefJ}
Davis T., Mortsell, Sullerman  et al., Scrutiniziting Exotic Cosmological Models using ESSENCE Supernova Data Combined with other Cosmological Probe [arXiv:astro-ph/0701510v2], (2007)
\bibitem{RefJ}
Deutsch D., The Structure of the Multiverse [arXiv:quant-ph/0104033], (2001)
\bibitem{RefJ}
Dirac P., Directions in Physics, New York, J. Wiley and Sons, 1978, Lecture 1, (1978)
\bibitem{RefJ}
Doplicher S., Piacitelly G., Tomassini L. and Vaggiu S., Comment on ``Can we measure structures to a precision better than the Planck length?'', by Sabine Hossenfelder [arXiv:gr-qc/1206.3067v1], (2012)
\bibitem{RefJ}
Durrer R. and Maartens R., Dark energy and dark gravity: theory overview, Gen. Rel. Grav. 40, 301 [arXiv:astro-ph/0711.0077v2], (2008)
\bibitem{RefJ}
Durrer R., What do we really know abut Dark Energy? [arXiv:astro-ph.CO:1103.5331v3], (2011)
\bibitem{RefJ}
Ellis G., Kirchner U., Stoger W., Multiverses and Physical Cosmology Mount. Not. Roy. Soc. 347, 921--936 [arXiv:astro-ph/0305292], (2004)
\bibitem{RefJ}
Ellis G., Does the Multiverse Really Exists? Sci. Am. 305, 38--43, (2011)
\bibitem{RefJ}
Ellis G., The Multiverse: conjecture, proof, and science. Talk at Nicolai Fest, Golm, (2012)
\bibitem{RefJ}
Ellis J., Unification and Super symmetry, Phil. Trans. R. Soc. Lond. A 310, 279, (1983)
\bibitem{RefJ}
Feynman R., Moringo F., Wagner W. and Hatfield B., Feynman Lectures on Physics. Vol.1 Ch. 24, Penguin, (1999)
\bibitem{RefJ}
Filippenko A., Evidence from Type Ia Supernovae for an Accelerating Universe and Dark Energy In Carnegie Observatories Astrophysics Series, Vol.2: Measuring and Modeling the Universe, ed. W. L. Freedman (Cambridge: Cambridge Univ. Press) [arXiv:astro-ph/0307139v1], (2003)
\bibitem{RefJ}
Filippenko A. and Riess A., Evidence from Type Ia Supernovae for an Accelerating Universe [astro-ph/0008057], (2000)
\bibitem{RefJ}
Fixsen D. and Mather J., The Spectral Results of the Far -- Infrared Absolute Spectrophotometer Instrument on COBE ApJ, 581, 817, (2002)
\bibitem{RefJ}
Freedman W., Madore B., Gibson B. and et al., Final Results from the Hubble Space Telescope Key Project to Measure the Hubble Constant, ApJ  553, 47, (2001)
\bibitem{RefJ}
Freundlich E., Red Shifts in the Spectra of Celestial Bodies, Phil. Mag., 45, 303, (1954)
\bibitem{RefJ}
Frieman J., Turner M. and Huterer D., Dark Energy and the Accelerating Universe, Ann. Rev. Astron. Astrophys, 46, 385 [arXiv:astro-ph/0803.0982], (2008)
\bibitem{RefJ}
Fritzsch H. and Sola J., Matter Non-conservation in the Universe and Dynamical Dark Energy [arXiv:hep-ph/1202.5097v2], (2012)
\bibitem{RefJ}
Garcia --Asperita M., A Geometric Solution for Dark Energy: Vanishing Dimension [arXiv:gr-qc/ 1109.5177v3], (2011)
\bibitem{RefJ}
Gasanalizade A., On the Weighted Factor f in the Temperature-Time Relation of the Early Universe. In Proc. Int. Conf. ``Solar System Research'', eds. M. Karimov \& A. Guliev (Baku: ``Elm''), p.16, (2004)
\bibitem{RefJ}
Gasanalizade A., The Application of the Freundlich Redshift Hypothesis to the Big Bang Cosmology: Predicting the CMB Temperature, Size, Age and Mass of the Universe, in Proc. IX Int. Conf. ``Space, Time, Gravitation'' held on July 7--11, 2006. Saint-Petersburg, Russia, eds. A. Aldoshin \& M. Varin, p. 322, ``Tessa'', St-Petersburg, (2007)
\bibitem{RefJ}
Gasanalizade A., On Synthesis of the Big Bang Model with Freundlich's Redshift and its Cosmological Consequences [arXiv:physics.gen-ph/ 1009.4775v3], (2010)
\bibitem{RefJ}
Gasanalizade A., On the Determination of Dark Energy and Dark Matter from Planck and Trans-Planck Redshifts [arXiv:physics.gen-ph/1111.2936v1], (2011)
\bibitem{RefJ}
Gell-Mann M., Ramond P. and Slanski R., in ``Supergravity'' eds. P. van Niuwenhuizen and D. Freedman, Proc. Supergravity Workshop at Stone Brook, Sept. 1979, p. 315, North-Holland Publ. Co., Amsterdam, (1979)
\bibitem{RefJ}
Glashow S., Particle Physics in the United States A Personal View [arXiv: hep-ph/1305.5482v1], (2013)
\bibitem{RefJ}
Gott R., Vogeley M., Podariu S. and Ratra B., Median Statistics $H_0$, and the Accelerating Universe, ApJ, 549, 1 [arXiv:astro-ph/0006103], (2001)
\bibitem{RefJ}
Gundlach J. and Merkowitz S., Measurement of Newton's Constant Using a Torsion Balance with Angular Acceleration Feedback, Phys. Rev. Lett. 85, 2869, (2000)
\bibitem{RefJ}
Guth A., The Inflationary Universe: A Possible Solution to the Horizon and Flatness Problems: Phys. Rev. D 23, 34, (1981)
\bibitem{RefJ}
Hansson J., On the Origin of Elementary Particle Masses, [arXiv:physics.gen-ph/1211.3136v1], (2012)
\bibitem{RefJ}
Haxton W., Cosmology and Fundamental Physics and their Laboratory Astrophysics Connections [arXiv:astro-ph/1101.2699v1], (2011)
\bibitem{RefJ}
Hinshaw G., Weiland J., Hill R. et al., Five-Year Wilkinson Microwave Anisotropy Probe (WMAP) Observations: Data Processing, Sky Maps, and Basic Results, ApJS, 180, 225 [arXiv:astro-ph.CO/0803.0732v2], (2009)
\bibitem{RefJ}
't Hooft G., Dimensional reduction in quantum gravity [arXiv:gr-qc/9310026v2], (2009)
\bibitem{RefJ}
Hossenfelder S., Can we measure structures to a precision better than the Planck length? Class. Quant. Grav. 29, 115011. [arXiv:1205.3636], (2012)
\bibitem{RefJ}
Jarosik N., Bennett C., Dunkey J. et al., Seven-Year Wilkinson Microwave Anisotropy Probe (WMAP) Observations: Sky Maps, Systematic Errors, and Basic Results, ApJS, 192, 14 [arXiv: astro-ph.CO/ 1001.4744v1], (2011)
\bibitem{RefJ}
Kittel C., Knight W. and Ruderman M., Mechanics. Berkeley Physics Course, [Ch. 9], Mc Graw -- Hill Book Company, (1964)
\bibitem{RefJ}
Knop R., Aldering G., Amanullah R. et al., (The Supernova Cosmology Project), New Constraints on $\Omega_M$, $\Omega_\Lambda$, and $w$ from an Independent Set of Eleven High-Redshift Supernovae Observed with HST, ApJ, 598, 102 [arXiv:astro-ph/0309368], (2003)
\bibitem{RefJ}
Kolb E., Starobinsky A. and Tkachev I., Trans-Planckian wimpzillas, JCAP, 0707, 005 [arXiv: hep-th/0702143v2], (2007)
\bibitem{RefJ}
Komatsu E., Dunkley J., Nolta M. et. al., Five-Year Wilkinson Microwave Anisotropy Probe (WMAP) Observations: Cosmological Interpretation, ApJS, 180, 330 [arXiv:astro-ph.CO/0803.0547v2], (2008)
\bibitem{RefJ}
Komatsu E., Smith K., Dunkley J. et al., Seven-Year Wilkinson Microwave Anisotropy Probe (WMAP) Observations: Cosmological Interpretation, ApJS, 192, 18 [arXiv:astro-ph. CO/1001.4538v3], (2011)
\bibitem{RefJ}
Kondepudi D. and Prigogine J., Modern Thermodynamics: From Heat Engines to Dissipative Structures, (Ch.5), John Wiley and Sons, New York, (1999) 
\bibitem{RefJ}
Koschmieder E., Theory of the Elementary Particles [arXiv:physics. gen-ph/0804.4848], (2013)
\bibitem{RefJ}
Kragh H., What's in a Name: History and Meanings of the Term ``Big Bang'' [arXiv:1301.0219v1], (2013)
\bibitem{RefJ}
Krauss L. and Dent J., A Higgs-Saw Mechanism as a Source for Dark Energy [arXiv:hep-ph/1306.3239v1], (2013)
\bibitem{RefJ}
Kroupa P., Pawlowski M. and Milgrom M., The Failures of the Standard model of cosmology require a new paradigm, [arXiv:1301.3907v1], (2013) 
\bibitem{RefJ}
Kuhne R., Time-Varying Fine-Structure Constant Requires Cosmological Constant, [arXiv:astro-ph/9908356v1], (1999)
\bibitem{RefJ}
Israelit M., Dark Energy in a perturbed Weil-Dirac Universe, [arXiv:1212.2025v1], (2012)
\bibitem{RefJ}
Lahav O. and Liddle A., The Cosmological Parameters [arXiv:astro-ph.CO/1002.3488v1], (2010)
\bibitem{RefJ}
Lampton M., Dark Energy--What is--What it means -- \\ 
(https://www.ssl.berkeley.edu/\textasciitilde mlampton/DarkEnergy.pdf), (2012)
\bibitem{RefJ}
Langacker P., in ``An advocate for the anthropic principle'' Review: in Physics Today, 59, June 6, p.61, (2006)
\bibitem{RefJ}
Larson D., Dunkley J,. Hinshaw G. et al., Seven-Year Wilkinson Microwave Anisotropy Probe (WMAP) Observations: Power Spectra and WMAP -- Derived Parameters, ApJS, 192, 16 [arXiv:astro-ph. CO/1001.4635v2], (2011)
\bibitem{RefJ}
Layzer D., Constructing the Universe (Scientific American Library), [Ch.1], An imprint of Sc. Am. Books, Inc., New York, (1984)
\bibitem{RefJ}
Lemson G. (VIRGO), Halo and Galaxy Formation Histories from the Millennium Simulations [arXiv:astro-ph/0608019], (2006)
\bibitem{RefJ}
Li Z., Wu P. and Yu H. 2010 Probing the course of cosmic expansion with a combination of observational data [arXiv:gr-qc/1011.2036v1]
\bibitem{RefJ}
Li Z., Wu P. and Yu H., Examining the cosmic acceleration with the latest Union 2 supernova data [arXiv: gr-qc/1011.1982v2], (2011)
\bibitem{RefJ}
Li Miao, Li X.-D., Wang S. and Wang Y., Dark Energy, Commun. Theor. Phys. 56, 525 [arXiv:astro-ph.CO/1103.5870v6], (2011)
\bibitem{RefJ}
Li Miao, Li X.-D., Wang S. and Wang Y., Dark Energy: Brief Review [arXiv:astro-ph. CO/1209.0922v1], (2012)
\bibitem{RefJ}
Linde A., Nonsingular Regenerating Inflationary Universe, (CUP Preprint-82-0554;\\ http://web.stanford.edu/\textasciitilde alinde/1982.pdf), (1982)
\bibitem{RefJ}
Linde A., The New Inflationary Universe Scenario, in The Very Early Universe, ed. By G. Gibbons, S. Hawking and S. Siklos, (CUP, Cambridge). pp. 205--249., (1983)
\bibitem{RefJ}
Linde A., Particle Physics and Inflationary Cosmology, (Chur: Harwood Academic Publshers), (1990)
\bibitem{RefJ}
Linde A. and Vanchurin V., How many universes are in the multiverse? [arXiv:hep-th/0910.1589v1], (2009)
\bibitem{RefJ}
Linder E., Frontiers of Dark Energy, [arXiv:astro-ph/1009.1411], (2010)
\bibitem{RefJ}
Lopez-Corredoira M., Observational Cosmology: caveats and open questions in the standard model, [arXiv:astro-ph/03102114v2], (2003)
\bibitem{RefJ}
Melvin M., Freundlich Red -- Shift Formula Phys. Rev., 98, 884, (1955)
\bibitem{RefJ}
Mersini L. Bastero-Gil M. and Kant P., Relic dark energy from the trans-Plankian regime, Phys. Rev. D 64, 043508, (2001)
\bibitem{RefJ}
Mohr P., Taylor B. and Newell D., CODATA Recommended Values of the Fundamental Physical  Constants: 2006, Rev. Mod. Phys., 80, 633 (http://physics.nist.gov/constants), (2008)
\bibitem{RefJ}
Mohr P. and Newell D., Resource Letter FC-1: The Physics of Fundamental Constants, Am. J. Phys. 78, 338, (2010)
\bibitem{RefJ}
Nguyen H., Scale invariance in cosmology and physics, [arXiv:physics.gen-ph /1111.5529v1], (2011)
\bibitem{RefJ}
Nicolay H., Quantum Gravity: The view from particle physics, [arXiv:gr-qc/1301.5481v1], (2013)
\bibitem{RefJ}
Padmanabhan T., Cosmological Constant -- the Weight of the Vacuum, Phys. Rept. 380, 235 [arXiv:hep-th/0212290v2], (2003)
\bibitem{RefJ}
Padmanabhan T., Dark Energy: Mystery of Millennium, AIP Conf. Proc. 861, 179 [arXiv:astro-ph/0603114], (2006)
\bibitem{RefJ}
Padmanabhan T., Emergent Gravity and Dark Energy [arXiv:gr-qc/0802.1798v1], (2008)
\bibitem{RefJ}
Padmanabhan T., Why Does the Universe Expand? Gen. Rel.Grav., 42, 2743 [arXiv:gr-qc/1001.3380v1], (2010)
\bibitem{RefJ}
Peebles P., Phenomenology of the Invisible Universe,[arXiv:astro-ph/0910.5142v1], (2009)
\bibitem{RefJ}
Peebles P., Dark Matter [arXiv:astro-ph.CO/1305.6859v1], (2013)
\bibitem{RefJ}
Peebles P. and Ratra B., The Cosmological Constant and Dark Energy, Rev. Mod. Phys. 75, 559, [arXiv:astro-ph/0207347v2], (2003)
\bibitem{RefJ}
Penrose R., The Singularity in Cosmology, IAU Symposium 63, ed. by M.S. Longair, Reidel, Dordrecht, (1974)
\bibitem{RefJ}
Penrose R., The Emperor's New Mind, [Chap. 8], OUP, Oxford, (1989)
\bibitem{RefJ}
Perlmutter S. Aldering G., Goldhaber G. et al., Measurements of $\Omega$ and $\Lambda$ from 42 High-Redshift Supernovae, ApJ, 517, 565 [arXiv:astro-ph/0005265], (1999)
\bibitem{RefJ}
Petrosian V., Lemaitre Model, Cosmological Constant and Observations, IAU Symposium 63, ed. by M.S. Longair, Reidel, Dordrecht, (1974)
\bibitem{RefJ}
Planck M., Ueber Irreversible Strahlungs vor gange, Ann. Phys., 1, 69, (1900)
\bibitem{RefJ}
Polarski D., Dark Energy: Beyond general relativity? AIP Conf. Proc. 861, 1013 [arXiv:astro-ph/0605532], (2006)
\bibitem{RefJ}
Ratra B. and Vogeley M., The Beginning and Evolution of the Universe, Publ. of the Astr. Soc. Pacific 120, 235 (March 2008),[arXiv:astro-ph/0706.1567v2], (2008)
\bibitem{RefJ}
Rees M., Our Cosmic Habitat, (Princeton: Princeton University Press) (Parts II and III), (2001)
\bibitem{RefJ}
Reynaud S., Lamine B. and Jackel M., In Proceedings of the Varenna scool on Atomic Optics and  Space Physics (Societa Italiane di Fisica, 2008), [arXiv:gr-qc/0801.3411v1], (2008)
\bibitem{RefJ}
Riess A., Filipenko A., Challis P. et al., Observational Evidence from Supernovae for an Accelerating Universe and a Cosmological Constant, Astron. J. 116, 1009 [arXiv:astro-ph/9805201], (1998)
\bibitem{RefJ}
Riess A., .Strolger L.-G, Tonry J. et al., Type Ia Supernova Discoveries at $z>1$ From the Hubble Space Telescope: Evidence for Past Deceleration and Constraints on Dark Energy Evolution, ApJ, 627, 579 [arXiv:astro-ph/0402512v2], (2005)
\bibitem{RefJ}
Sahni V. and Starobinsky A., Reconstructing Dark Energy, Int. J. Mod. Phys. D15, 2105 [arXiv:astro-ph/0610026], (2006)
\bibitem{RefJ}
Sandage A.,Tamman A., Saha A., Reindl B., Macchetto F. and Panagia N., The Hubble Constant: A summary of the Hubble Space Telescope program for the Luminosity of Type Ia Supernovae by means of Cepheids'', ApJ, 653, 843, (2006)
\bibitem{RefJ}
Schramm D., The early universe and high-energy physics, Physics Today, April 4, p. 27, (1983)
\bibitem{RefJ}
Silk J, The Big Bang. The creation and evolution of the Universe., [Ch.17], W. H. Freeman and Company, San Francisco, (1980)
\bibitem{RefJ}
Singal J., Fixsen D., Kogut A., Levin S., Limon M., Mirel P., Seiffert M. and Wollak E., The Cosmic Microvawe Background Temperature and Galactic Emission at 8.0 and 8.3 GHz, ApJ, 653, 835, (2006)
\bibitem{RefJ}
Spergel D., Bean R., Dore O. et al., Wilkinson Microwave Anisotropy Probe (WMAP) Three Year Observations: Implications for Cosmology [WMAP Collaboration], ApJ, 657, 645 [arXiv:astro-ph/0603449v2], (2007)
\bibitem{RefJ}
Steinhardt P. and Turok N., A Cyclic Model of the Universe, Science 296, 1436, (2002a)
\bibitem{RefJ}
Steinhardt P. and Turok N., Cosmic Evolution in a Cyclic Universe, Phys. Rev. D 65, 126003, (2002b)
\bibitem{RefJ}
Steinhardt P. and Turok N., The Cyclic Model Simplified [arXiv:astro-ph/0404480v1], (2004)
\bibitem{RefJ}
Stoeger W., Ellis G. and Kirchner U., Multiverse and Cosmology: Philosophical Issues [arXiv:astro-ph/0407329v2], (2004)
\bibitem{RefJ}
Stoeger W., Retroduction, Multiverse Hypotheses and Their Testability [arXiv:astro-ph/0602356v2], (2006)
\bibitem{RefJ}
Straumann N., Dark Energy: Recent Developments, Mod. Phys. Lett., A21, 1083 [arXiv: hep-ph/0604231v1], (2006)
\bibitem{RefJ}
Sullivan M, Guy J, Conle A et al, SNLS3: Constraints on Dark Energy Combining the Supernova Legacy Survey Three Year Data with Other Probes [arXiv:astro-ph.CO/1104.1444v2], (2011)
\bibitem{RefJ}
Susskind L., The Cosmic Landscape: String Theory and the Illusion of Intelligent Design, Little, Brown and Co, New York. 2006 (Reviewed by P.Langacker in ``Phys. Today'' 59, Number 6, p. 61), (2006)
\bibitem{RefJ}
Suzuki N., Rubin D., Lideman C. et al., (The Supernova Cosmological Project) The Hubble Space Telescope Cluster Supernova Survey: V. Improving the Dark Energy Constraints above $z > 1$ and Build in an Early-Type -- Hosted Supernova Sample [arXiv:astro-ph. Co/1105.3470v1], (2011)
\bibitem{RefJ}
Suyu S., Treu T., Blanford R. et al., The Hubble constant and new discoveries in cosmology [arXiv:astro-ph/1202.4459v1], (2012)
\bibitem{RefJ}
Tegmark M., Parallel Universes Sci. Am. 40, 41 [arXiv:astro-ph/0302131], (2003)
\bibitem{RefJ}
Tegmark M., Aguirre A., Rees M. and Wilczek F., Dimensionless Constants, Cosmology, and Other Dark Matters Phys. Rev. D73, 023505 [arXiv:astro-ph/0511774], (2006)
\bibitem{RefJ}
Tegmark M., Eisenstein D., Strauss M. et al., Cosmological Constraints from the SDSS Luminous Red Galaxies Phys. Rev. D 74, 123507 [arXiv:astro-ph/0608632v2], (2006)
\bibitem{RefJ}
Thorne K., Black Holes and Time Warps. Einstein's Outrageous Legacy, [Chs. 12, 14], W. Norton and Company, NY, London, (1994)
\bibitem{RefJ}
Turner M., Dark Energy: Just what Theorists Ordered?, Physics Today, April 4, p. 10, (2003)
\bibitem{RefJ}
Turner M., Dark Energy, Particle Physics and Cosmology. Am. Astr. Soc., AAS Meeting \#220, \#316.01 [Bibliogrphic Code: 3012AAS…220316011], (2012)
\bibitem{RefJ}
Turner M. and Huterer D., Cosmic Acceleration, Dark Energy and Fundamental Physics, J. Phys. Soc. Jap., 76, 111015 [arXiv:astro-ph/0706.2186v2], (2007)
\bibitem{RefJ}
Turok N. and Steinhardt B., Beyond Inflation: A cyclic Universe Scenario, Phys. Scr., T117, a00076, (2005)
\bibitem{RefJ}
Turyshev S., Israelsson U., Shao M. et al., Space-based research in fundamental physics and quantum techonologies, Int. J. Mod. Phys,. D 16, 1879 [arXiv:gr-qc/0711.0150v3], (2007)
\bibitem{RefJ}
Vaas R., Dark Energy and Life's Ultimate Future In: Burdyuzha, Vladimir (Ed.): The Future of Life and the Future of our Civilization Springer: Dordrecht pp. 231--247. ISBN 978-1-4020-4967-5, (2006)
\bibitem{RefJ}
Verschuur G., A Close Look at the Relationship between WMAP (ILC) Small-scale Features and Galactic HI Structure Am. Astr. Soc., AAS Meeting \#220, \#504.02, (2012)
\bibitem{RefJ}
Verlinde E., On the Origin of Gravity and the Laws of Newton, JHEP 1104, 029 [arXiv:hep-th/1001.0785v1], (2010)
\bibitem{RefJ}
Weinberg S., The First Three Minutes. Basic Books Library; [Ch. 7], Andre Deutsch, New York, (1977)
\bibitem{RefJ}
Weinberg S., Dreams of a Final Theory. Vintage Books. A Division of Random House Inc., New York, (1992)
\bibitem{RefJ}
Weinberg S., Living in the Multiverse In Carr B (ed). Universe or Multiverse? (CUP, Cambridge) 29--42 [arXiv:hep-th/0511037v1], (2007)
\bibitem{RefJ}
Weyl H., Gravitation and Electricity, Sitzungsber d. Berl. Akad. p. 465, (1918)
\bibitem{RefJ}
White S., Fundamental Physics: why Dark Energy is bad for Astronomy, Rept. Prog. Phys., 70, 883-898 [arXiv:astro-ph/0704.2291v1], (2007)
\bibitem{RefJ}
Will C., The Confrontation between General Relativity and Experiment, Living Rev. Relativity, 9, 3 [arXiv:gr-qc/0510072v3], (2006)
\bibitem{RefJ}
Yoneya T., String Theory and the Space-Time Uncertainty Principle [arXiv:hep-th/0004074v6], (2001)
\bibitem{RefJ}
Zeldovich Y., Cosmological constant and elementary particles, Sov. Phys. JETP Lett., 6, 316, (1967)


\end{thebibliography}


\end{document}